\documentclass[fleqn,usenatbib]{mnras}

\usepackage{newtxtext,newtxmath}

\usepackage[T1]{fontenc}

\DeclareRobustCommand{\VAN}[3]{#2}
\let\VANthebibliography\thebibliography
\def\thebibliography{\DeclareRobustCommand{\VAN}[3]{##3}\VANthebibliography}

\usepackage{graphicx}	
\usepackage{amsmath}	

\title[Asteroseismic analysis of red giants]{Asteroseismic analysis of red giants in eclipsing binaries using two methods: implications for scaling relations and chemical composition}

\author[M. Y{\i}ld{\i}z, S. \"Ortel and T. \c{C}ak{\i}r Alsa\c{c}] {M. Y{\i}ld{\i}z$^1$$^{}$\thanks{E-mail:
mutlu.yildiz@ege.edu.tr}, S. \"Ortel$^2$ and T. \c{C}ak{\i}r Alsa\c{c}$^3$\\
$^1$Department of Astronomy and Space Sciences, Faculty of Science, Ege University, 35100 \.Izmir, Turkey.\\
$^2$Independent Researcher, Şişli 34377 İstanbul, Turkey\\
$^3$Department of Astronomy and Space Sciences, Graduate School of Natural and Applied Sciences, Ege University, 35100 \.Izmir, Turkey.}

\date{Accepted XXX. Received YYY; in original form ZZZ}

\pubyear{2024}

\begin{document}

\label{firstpage}
\pagerange{\pageref{firstpage}--\pageref{lastpage}}
\def\braket#1{\left<#1\right>}
\newcommand{\yildiz}{Y\i ld\i z }
\newcommand{\etal}{et al. }
\newcommand{\wrt}{with respect to }
\newcommand{\logg}{\log(g) }
\newcommand{\numino}{\mbox{\ifmmode{\overline{\nu_{\rm min}}}\else$\overline{\nu_{\rm min}}$\fi}}
\newcommand{\numin}{\mbox{\ifmmode{\nu_{\rm min}}\else$\nu_{\rm min}$\fi}}
\newcommand{\teff}{\mbox{\ifmmode{T_{\rm eff}}\else$T_{\rm eff}$\fi}}
\newcommand{\teffsun}{\mbox{\ifmmode{{\rm T}_{\rm eff{\sun}}}\else${\rm T}_{\rm eff{\sun}}$\fi}}
\newcommand{\numax}{\mbox{$\nu_{\rm max}$}}
\newcommand{\nuH}{\mbox{\ifmmode{\nu_{\rm minH}}\else$\nu_{\rm minH}$\fi}}
\newcommand{\nuL}{\mbox{\ifmmode{\nu_{\rm minL}}\else$\nu_{\rm minL}$\fi}}
\newcommand{\Dnu}{\mbox{$\Delta \nu$}}
\newcommand{\Dpi}{\mbox{$\Delta \upi$}}
\newcommand{\muHz}{\mbox{$\mu$Hz}}
\newcommand{\kepler}{\mbox{{\it Kepler}}}
\newcommand{\corot}{\mbox{{\it CoRoT}}}
\newcommand{\tess}{\mbox{{\it TESS}}}
\newcommand{\gaia}{\mbox{{\it Gaia}}}
\newcommand{\numaxS}{\mbox{$\nu_{\rm max {\sun}}$}}
\newcommand{\MS}{{\rm M}\ifmmode_{\sun}\else$_{\sun}$~\fi}
\newcommand{\RS}{{\rm R}\ifmmode_{\sun}\else$_{\sun}$~\fi}
\newcommand{\LS}{{\rm L}\ifmmode_{\sun}\else$_{\sun}$~\fi}
\newcommand{\MSbit}{{\rm M}\ifmmode_{\sun}\else$_{\sun}$\fi}
\newcommand{\RSbit}{{\rm R}\ifmmode_{\sun}\else$_{\sun}$\fi}
\newcommand{\LSbit}{{\rm L}\ifmmode_{\sun}\else$_{\sun}$\fi}
\maketitle

\begin{abstract}
{ The study of solar-like oscillating red giants in eclipsing binaries (EBs) provides a unique opportunity to advance stellar astrophysics by combining dynamical mass and radius measurements with asteroseismic constraints. EBs provide precise fundamental parameters (e.g. mass, radius, and luminosity) independent of distance, while solar-like oscillations probe stellar interiors and enable tests of asteroseismic scaling relations used to determine stellar masses and radii. {We apply two different methods to estimate the initial chemical composition of the systems. In Method I, the initial helium abundance ($Y_0$) is treated as the free parameter, whereas in Method II the free parameter is the initial metallicity ($Z_0$), assuming a relation between $Y_0$ and $Z_0$. We construct interior models individually for the components of 11 EBs and obtain coeval solutions for eight systems.} The ages and chemical compositions derived from the two methods are generally consistent with each other. Our results provide important clues about the chemical evolution of a part of the Galactic disk. Moreover, using the parameters obtained for two oscillating stars, Tek Ayak (KIC 8410637) and KIC 9970396, instead of solar reference values in the scaling relations yields masses and radii that are in much better agreement with the dynamical solutions without requiring additional corrections.  }
\end{abstract}




\section{Introduction}


Eclipsing binaries (EBs) in general, and more recently solar-like oscillating (SLO) stars in great, are of particular importance in obtaining fundamental stellar parameters and testing the theory of stellar evolution. EBs with SLO components are invaluable in this respect.
Such EB stars are among the Kepler targets \citep{2013A&A...556A.138F, 2016ApJ...832..121G, 2022A&A...667A..31B, 2024A&A...682A...7B, 2025arXiv250109018G}. The red giant (RG) component of some of these EBs exhibits solar-like oscillations. These stars should be considered very special laboratories for testing the theory of stellar evolution. Considering that most of the stars detected by the Transiting Exoplanet Survey Satellite \citep[{\it TESS},][]{Sullivan2015} and {\it Kepler} \citep{Borucki2010} missions are RGs, the importance of these SLO EBs becomes even clearer.

Joint binary star and { asteroseismic} modelling improves stellar age determination. In this regard, comprehensive studies have been conducted {recently by analysing high-quality data }\citep{2013A&A...556A.138F, 2016ApJ...832..121G, 2018MNRAS.476.3729B,2018MNRAS.478.4669T,2019MNRAS.484..451H,2022MNRAS.517.4187T, 2024A&A...682A...7B, 2025A&A...699A.152T, 2025arXiv250109018G}.
The mass ($M$), radius ($R$), effective temperature (\teff)  and metallicity ([Fe/H]) parameters of the components have been determined from observational spectral and photometric data \citep{2016ApJ...832..121G}. It is important to determine the ages and chemical compositions of the binary stars well for many reasons. For example, this information can shed light on the chemical evolution of the Milky Way. We can construct internal structure models of the components using common age and chemical composition constraints. These stars are very important in two respects. { First, because the evolutionary properties of the components are very different from each other, with the primary component being an RG and the secondary component being either a main-sequence star or close to the main sequence (MS) in nine systems \citep[see][]{2024arXiv240902983R}, we can determine the properties of the system uniquely.} These systems are perhaps the most suitable for a unique solution. Secondly, we can use {\ the oscillating RG component as a reference star} for the scaling relations between the asteroseismic and non-asteroseismic parameters of these stars. Thus, we use the values for this star instead of the solar values. The selected reference star can be used to calculate the fundamental parameters of stars with high-quality asteroseismic data \citep[e.g., RGs in APOKASC-2,][]{2018ApJS..239...32P}. For example, it is better to use an RG as the reference star instead of the Sun in scaling relations (see {Section \ref{sec:461}}).

The frequency at the maximum amplitude of the oscillations ($\nu_{\rm max}$) and the large frequency separation (${\Delta \nu}$) are very effective asteroseismic quantities to obtain fundamental properties of stars. { $\nu_{\rm max}$ approximately scales with 
$g/\sqrt{T_{\rm eff}}$, where $g$ represents gravity \citep{Brown1991,Kjebed1995}.} 
${\Delta \nu}$ is the average frequency separation between the oscillation modes with the same spherical degree $\ell$, but consecutive radial orders $n$. { The mean of $\Dnu$ ($\braket{\Dnu}$) approximately scales with} the square root of the mean density ($\braket{\rho}$) of the star \citep{1980ApJS...43..469T,1993ASPC...42..347C}. 
$\braket{\Dnu}$ and $\nu_{\rm max}$ are 
effective tools for computing the mass and radius of the oscillating stars. The standard scaling relations for the radius ($R_{\rm sca}$) and mass ($M_{\rm sca}$) of an oscillating star are given as:
\begin{equation}
\label{eq:clasicsca}
\frac{R_{\rm sca}}{\rm R_{\odot}}=\frac{\numax/\nu_{\rm max\odot}}{(\braket{\Delta \nu}/\braket{\Delta \nu_\odot})^2}\left( \frac{T_{\rm eff}}{\rm T_{\rm eff\odot}}
\right)^{1/2},
\end{equation}

\begin{equation}
\frac{M_{\rm sca}}{\rm M_{\odot}}=\frac{(\numax/\nu_{\rm max\odot})^3}{(\braket{\Delta \nu}/\braket{\Delta \nu_\odot})^4}\left( \frac{T_{\rm eff}}{\rm T_{\rm eff\odot}}
\right)^{3/2}, \nonumber
\end{equation}
respectively. Here, ${\braket{\Dnu_{\sun}}}$ and $\nu_{\rm max\odot}$ are the solar values. 
These values are as follows: ${\braket{\Dnu_{\sun}}}=135.1$ $\mu$Hz and $\nu_{\rm max\odot}=3090$ $\mu$Hz \citep{Sharma}. 

{The non-standard scaling relations \citep{Sharma,2017ApJ...844..102H,2022ApJ...927..167L,2023MNRAS.518.5552Y} predict $M$ and $R$ of the SLO components more realistically than the above-mentioned standard relations.} In these relations, the correction parameters $f_{\nu_{\rm max}}$ and $f_{\Delta \nu}$ are used to correct the scaling relations ${\nu_{\rm max}}$ and ${\Delta \nu}$, respectively. 
{To derive the non-standard scaling relations for radius ($R'_{\rm sca}$), 
        $R_{\rm sca}$ must be multiplied by $f_{\Delta \nu}^2/f_{\nu_{\rm max}}$ as shown below:
                \begin{equation}
                \label{eq:non-standardRsca}
        \frac{R'_{\rm sca}}{\rm R_{\odot}}=\frac{(\numax/\nu_{\rm max\odot})}{(\braket{\Delta \nu}/\braket{\Delta \nu_\odot})^2}\left( \frac{T_{\rm eff}}{\rm T_{\rm eff\odot}}
                        \right)^{1/2}
        \frac{f_{\Delta \nu}^2}{f_{\nu_{\rm max}}}.
        \end{equation}
        For the non-standard scaling relation for mass ($M'_{\rm sca}$), $M_{\rm sca}$ should be multiplied by $f_{\Delta \nu}^4/f_{\nu_{\rm max}}^3$:
                \begin{equation}
                \label{eq:non-standardMsca}
        \frac{M'_{\rm sca}}{\rm M_{\odot}}=\frac{(\numax/\nu_{\rm max\odot})^3}{(\braket{\Delta \nu}/\braket{\Delta \nu_\odot})^4}\left( \frac{T_{\rm eff}}{\rm T_{\rm eff\odot}}
                        \right)^{3/2}
        \frac{f_{\Delta \nu}^4}{f_{\nu_{\rm max}}^3}.
                \end{equation}
}
The standard scaling relations assume that the values of $f_{\nu_{\rm max}}$ and $f_{\Delta \nu}$ are unity. Here, $f_{\Delta \nu}$ can be obtained from the interior models \citep{Sharma}, while $f_{\nu_{\rm max}}$ can be determined by comparing $R_{\rm sca}$ and
the radius ($R_{\pi}$) computed from the parallaxes. Using data from the APOKASC-2 catalogue and GAIA DR2 data \citep{Gaia2018}, \cite{2021MNRAS.504.2273Y} found that the mean value of 
 $f_{\nu_{\rm max}}$ was 1.003 for the K-mag data of 2MASS \citep{2006AJ....131.1163S}. However, \cite{2023MNRAS.518.5552Y} 
 proposed a new parametrisation of $f_{\nu_{\rm max}}$  for RGs and found that $f_{\nu_{\rm max}}$   depends on metallicity, ${\Delta \nu}$ and effective temperature. This approach is more successful than the method used by \cite{2021MNRAS.504.2273Y} to predict the $M$ and $R$ of the RGs. 
 
{The mass$-$metallicity (MZ) diagram is particularly important in determining the structural and evolutionary status of red giant branch (RGB) stars, particularly for mass-gaining and mass-losing processes. Most RGBs in APOKASC-2 are populated in a triangle in the MZ diagram \citep{2023MNRAS.518.5552Y}.} The outsiders are potential stars that have gained or lost mass. The position of the 11 SLO components in the diagram should be considered when constructing the interior models of these stars {(see Section \ref{sec:11SLOMZ})}.


{ In this study, we investigate 11 EB systems containing SLO RG components. Our main aim is to determine the ages and initial chemical compositions of these systems by constructing stellar interior models constrained by both dynamical and asteroseismic parameters. We apply two different modelling methods based on different assumptions for the initial chemical composition and compare the resulting stellar parameters, ages, and asteroseismic properties. We also examine the applicability of the asteroseismic scaling relations for RGs in EBs and investigate whether some well-constrained SLO stars can be used as alternative reference stars instead of the Sun.}

{The rest of this} paper is organised as follows. Section \ref{sec:sec2} presents and discusses the observational data of the binaries. The models are described in Section \ref{sec:sec3}. Section \ref{sec:sec4} presents and discusses the results.  Finally, Section \ref{sec:sec5} presents the conclusions of the study. {Computational method for age of the RGs from the grids are given in the Appendix \ref{sec:AppA}. { Appendix B is devoted to the individual notes on the binaries.}}

\section{Observational data and their implications for stellar structure and evolution}
The observational basic parameters of the 11 EB stars were {compiled from the literature} \citep{2013A&A...556A.138F, 2016ApJ...818..108R,2016ApJ...832..121G,2018MNRAS.476.3729B, 2018MNRAS.478.4669T, 2019MNRAS.484..451H,2022A&A...668A..82B,2022MNRAS.517.4187T, 2024A&A...682A...7B, 2025A&A...699A.152T}. 
{The primary stars (oscillating components) are all located in the region of the RGs. 
{These stars may be ascending stars in the RGB or core helium burning (CHeB, red clump (RC) or secondary clump) stars  (see Section \ref{sec:sec4}). }Most of the secondary components are around the MS. One of the secondary components is about to reach the RGB (KIC 4054905 B), while another appears to be slightly more evolved (KIC 9246715 B).}
\label{sec:sec2}

\subsection{Basic properties of the component stars from binary dynamics and spectra}
\begin{figure}
    \centering
        \includegraphics[width=1.2\linewidth]{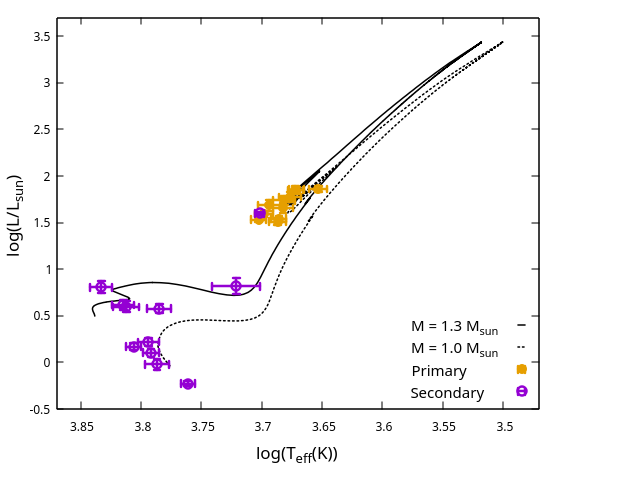}
    \caption{Hertzsprung$-$Russell diagram (HRD) for the components of the SLO EBs. Filled circles indicate the SLO primaries, while circles indicate the secondary components. The solid and dashed lines show the evolution path from the MS to the RG for the 1.3 and 1.0 M$_{\sun}$ models, respectively. For these models, the values $Y=0.2671$, $Z =0.01$ and, $\alpha=1.8311$ are assumed. Most of the secondary components are in the MS region, while one is in the RG region, and the other is very close to this region. The primary components are concentrated around the RG. 
    }
    \label{fig:1}
\end{figure}

{ In EB light curves, the depths of the eclipses provide mainly information about the { relative temperatures and surface brightnesses of the components, while the eclipse widths and shapes, together with the orbital geometry and semi-major axis, constrain the radii of the components.} The radial velocity curves obtained from the spectroscopic observations also contain information on the masses of the components. These parameters are compiled from the literature and are listed in Table \ref{tab:EBobs}. References are cited in the last column of the table. {We found results from 19 analyses for the 11 binaries.} The components of these binaries are plotted in the Hertzsprung$-$Russell diagram (HRD, Fig. \ref{fig:1}). Although most of the secondary (non-oscillating) components are located around the MS phase, the oscillating RG components have luminosity ($L$) values around $\log(L/{\rm L_{\odot}})=1.5-2.0$ and $R=7.48-14.1$ $\rm R_{\odot}$.}
\begin{table*}
\small\addtolength{\tabcolsep}{-3pt}
  \centering
 \caption{Observational physical properties of the EBs with SLO components. The dynamical values of KIC 4663623, KIC 5786154, KIC 7037405, KIC 7377422, KIC 8430105, KIC 9540226, and KIC 100001167 are taken from \protect\cite{2016ApJ...832..121G}. Ref. No.: 1 - \protect\cite{2022A&A...668A..82B}, 2- \protect\cite{2016ApJ...832..121G}, 3- \protect\cite{2018MNRAS.476.3729B}, 4- \protect\cite{2013A&A...556A.138F}, 5- \protect\cite{2018MNRAS.478.4669T}, 6- \protect\cite{2016ApJ...818..108R}, 7- \protect\cite{2021A&A...648A.113B}, 8- \protect\cite{2022MNRAS.517.4187T}, 9- \protect\cite{2019MNRAS.484..451H} and 10- \protect\cite{2025A&A...699A.152T}. For seven EBs, more than one measurement is available in the literature. The reference numbers in bold indicate the studies from which parameters are taken for modelling the component stars. {For the four systems (KIC 8410637, KIC 8430105, KIC 9540226, and KIC 10001167), the models are constructed for two observational data sets.} For KIC 8410637 (Tek Ayak), we also construct models (Model 8410637TzT in Table \ref{tab:Yol12_t9}) for which $T_{\rm effB}$=6380 K \citep{2022MNRAS.517.4187T}. The columns sequentially list KIC ID, mass and radius of the primary component ($M_{\rm A}$ and $R_{\rm A}$), mass and radius of the secondary component ($M_{\rm B}$ and $R_{\rm B}$), effective temperature of the primary and secondary components ($T_{\rm effA}$ and $T_{\rm effB}$), orbital eccentricity (e), period (P), metallicity ([Fe/H]), surface metal abundance ($Z_{\rm s}$), and reference (Ref).}
    \begin{tabular}{cllllllrrrrr}
    \hline
    \multicolumn{1}{l}{ KIC    } & \multicolumn{1}{c}{$M_{\rm A}$} & \multicolumn{1}{c}{$R_{\rm A}$ } & \multicolumn{1}{c}{$M_{\rm B}$} & \multicolumn{1}{c}{$R_{\rm B}$ } & \multicolumn{1}{c}{$T_{\rm effA}$} & \multicolumn{1}{c}{$T_{\rm effB}$} & \multicolumn{1}{c}{e} & \multicolumn{1}{c}{P} & \multicolumn{1}{c}{[Fe/H]} & \multicolumn{1}{c}{$Z_{\rm s}$} & \multicolumn{1}{r}{Ref} \\
    \multicolumn{1}{c}{} & \multicolumn{1}{c}{($\rm M_{\odot}$)} & \multicolumn{1}{c}{($\rm R_{\odot}$)} & \multicolumn{1}{c}{($\rm M_{\odot}$)} & \multicolumn{1}{c}{($\rm R_{\odot}$)} & \multicolumn{1}{c}{(K)} & \multicolumn{1}{c}{(K)} & \multicolumn{1}{c}{} & \multicolumn{1}{c}{(d)} & \multicolumn{1}{c}{} & \multicolumn{1}{c}{} & \multicolumn{1}{r}{} \\
    \hline
    4054905  & 0.954 $\pm$ 0.009 & 8.364  $\pm$ 0.027 & 0.957 $\pm$ 0.006 & 3.091 $\pm$ 0.009 & 4850 $\pm$ 70  & 5260 $\pm$ 240 & 0.37201 & 274.7288 & -0.60 $\pm$ 0.02 & 0.0034 $\pm$ 0.0002 & {\bf 1 }\\
    {"}      & 0.95 $\pm$ 0.04 & 8.19  $\pm$ 0.08 & 0.93 $\pm$ 0.01 & 3.11 $\pm$ 0.03 & 4790 $\pm$ 190  & 5100 $\pm$ 197 & 0.372 & 274.7306 & -0.71 $\pm$ 0.31 & 0.0026 $\pm$ 0.0013 & 7 \\
    4663623  & 1.36   $\pm$ 0.09   & 9.7    $\pm$ 0.2   & 1.34   $\pm$ 0.07   & 1.82   $\pm$ 0.06   & 4812 $\pm$ 92  & 6808 $\pm$ 140 & 0.43    & 358.0900 & -0.13 $\pm$ 0.06 & 0.0099 $\pm$ 0.0013 & 2 \\
    5786154  & 1.06   $\pm$ 0.06   & 11.4   $\pm$ 0.2   & 1.02   $\pm$ 0.04   & 1.59   $\pm$ 0.03   & 4747 $\pm$ 100 & 6527 $\pm$ 138 & 0.3764  & 197.9180 & -0.06 $\pm$ 0.06 & 0.0117 $\pm$ 0.0015 & {\bf 2} \\
    7037405  & 1.25   $\pm$ 0.04   & 14.1   $\pm$ 0.2   & 1.14   $\pm$ 0.02   & 1.80   $\pm$ 0.02   & 4516 $\pm$ 36  & 6303 $\pm$ 53  & 0.238   & 207.1083 & -0.34 $\pm$ 0.01 & 0.0061 $\pm$ 0.0001 & 2 \\
    {"}      & 1.17   $\pm$ 0.02   & 14.00 $\pm$ 0.09 & 1.11  $\pm$ 0.01  & 1.75  $\pm$ 0.01  & 4500 $\pm$ 80  & 6094 $\pm$ 138 & 0.2364  & 207.1085 & -0.27 $\pm$ 0.10 & 0.0072 $\pm$ 0.0015 & {\bf 3 }\\
    7377422  & 1.05   $\pm$ 0.08   & 9.5    $\pm$ 0.2   & 0.85   $\pm$ 0.03   & 0.87   $\pm$ 0.02   & 4938 $\pm$ 110 & 6120 $\pm$ 143 & 0.4377  & 107.6213 & -0.33 $\pm$ 0.06 & 0.0063 $\pm$ 0.0008 & {\bf 2} \\
    8410637  & 1.56   $\pm$ 0.03   & 10.7   $\pm$ 0.1   & 1.32   $\pm$ 0.02   & 1.57   $\pm$ 0.03   & 4800 $\pm$ 100 & 6490 $\pm$ 160 & 0.6864  & 408.3241 &  0.16 $\pm$ 0.03 & 0.0194 $\pm$ 0.0013 & {\bf 4} \\
    {8410637T}      & 1.47  $\pm$ 0.02  & 10.60 $\pm$ 0.05 & 1.31  $\pm$ 0.01  & 1.56  $\pm$ 0.01   & 4605 $\pm$ 80  & 6066 $\pm$ 200 & 0.686   & 408.3248 &  0.02 $\pm$ 0.08 & 0.0140 $\pm$ 0.0028 & {\bf 5} \\
    8430105  & 1.31   $\pm$ 0.02   & 7.65   $\pm$ 0.05  & 0.83   $\pm$ 0.01   & 0.770  $\pm$ 0.005  & 5042 $\pm$ 68  & 5771 $\pm$ 78  & 0.2564  & 63.32713 & -0.49 $\pm$ 0.04 & 0.0043 $\pm$ 0.0004 & {\bf 2} \\
    {8430105T}      & 1.25 $\pm$ 0.01 & 7.48 $\pm$ 0.03 & 0.813 $\pm$ 0.005 & 0.749 $\pm$ 0.004 & 4990 $\pm$ 80 & 5655 $\pm$ 80 & 0.25644 & 63.327106 & -0.46 $\pm$ 0.10 & 0.0046 $\pm$ 0.0012& {\bf 8}\\
    9246715  & 2.171  $\pm$ 0.007  & 8.37   $\pm$ 0.05 & 2.149  $\pm$ 0.007  & 8.30   $\pm$ 0.04 & 4990 $\pm$ 90  & 5030 $\pm$ 45  & 0.3559  & 171.2769 &  0.05 $\pm$ 0.02 & 0.0150 $\pm$ 0.0007 & 6 \\
    {"}      &  2.187 $\pm$ 0.003 & 8.49 $\pm$ 0.12 & 2.160 $\pm$ 0.003 & 8.20 $\pm$ 0.09 & 4890 $\pm$ 50 & 4905 $\pm$ 60 & 0.3552 & 171.2770 & 0.01 $\pm$ 0.03 & 0.0137 $\pm$ 0.0009 & {9}\\
    9540226  & 1.33   $\pm$ 0.05   & 12.8   $\pm$ 0.1   & 0.98   $\pm$ 0.03   & 0.99   $\pm$ 0.01   & 4692 $\pm$ 65  & 6399 $\pm$ 90  & 0.3880  & 175.4439 & -0.33 $\pm$ 0.04 & 0.0063 $\pm$ 0.0006 & {\bf 2 }\\
    {"}      & 1.39  $\pm$ 0.03  & 13.43  $\pm$ 0.17  & 1.02  $\pm$ 0.02  & 1.03  $\pm$ 0.01  & 4585 $\pm$ 75  & 5822 $\pm$ 200 & 0.3877  & 175.4438 & -0.31 $\pm$ 0.09 & 0.0066 $\pm$ 0.0012 & 5 \\
    {{9540226B}}      & 1.38  $\pm$ 0.04  & 13.06  $\pm$ 0.16  & 1.00  $\pm$ 0.02  & 1.01  $\pm$ 0.01  & 4680 $\pm$ 80  & 6157 $\pm$ 131 & 0.38782 & 175.443  & -0.23 $\pm$ 0.10 & 0.0079 $\pm$ 0.0016 & {\bf{3}} \\
    9970396  & 1.18  $\pm$ 0.02  & 8.04  $\pm$ 0.07 & 1.003 $\pm$ 0.009 & 1.109 $\pm$ 0.005 & 4860 $\pm$ 80  & 6221 $\pm$ 125 & 0.1942  & 235.2986 & -0.20 $\pm$ 0.02 & 0.0085 $\pm$ 0.0004 & {\bf 3} \\
    10001167 & 0.81   $\pm$ 0.05   & 12.7   $\pm$ 0.3   & 0.79   $\pm$ 0.03   & 0.98   $\pm$ 0.02   & 4700 $\pm$ 66  & 6191 $\pm$ 91  & 0.159   & 120.3903 & -0.69 $\pm$ 0.04 & 0.0027 $\pm$ 0.0002 & {\bf 2} \\
    {10001167T}      &  0.934 $\pm$ 0.008 & 13.03 $\pm$ 0.12 & 0.839$\pm$ 0.004 & 0.995 $\pm$ 0.015 & 4625 $\pm$ 29 & 6031 $\pm$ 108 & 0.15699 & 120.39005 & -0.73 $\pm$ 0.10 & 0.0025 $\pm$ 0.0006 & {\bf{10}}\\ 
     \hline
    \end{tabular}
  \label{tab:EBobs}
\end{table*}
\begin{figure}
    \centering
            \includegraphics[width=1.2\linewidth]{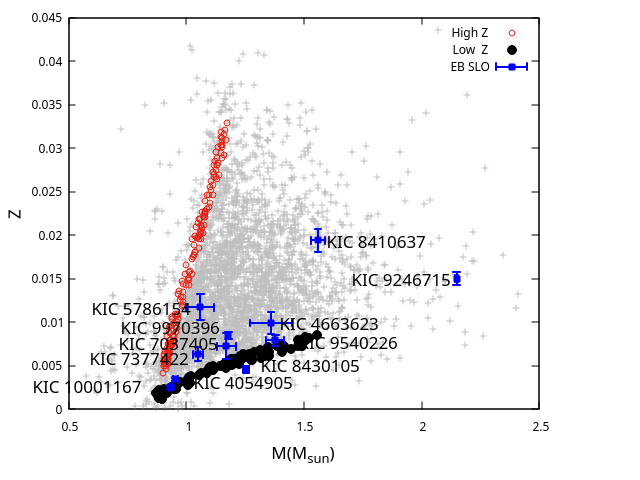}
            \includegraphics[width=1.2\linewidth]{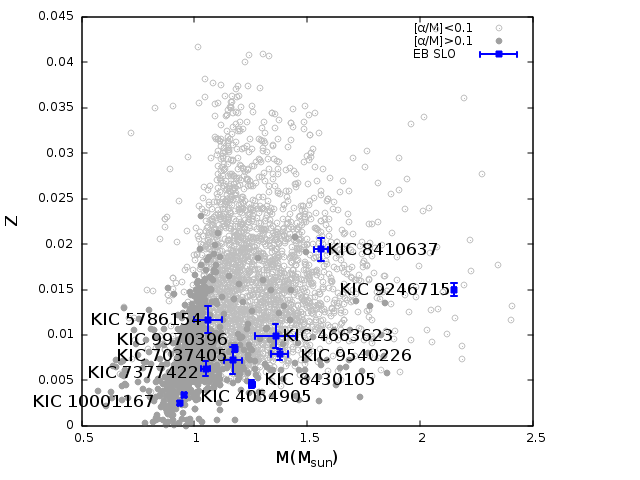}
    \caption{MZ diagram for 11 SLO stars in EBs. The crosses show the masses calculated by \protect\cite{2023MNRAS.518.5552Y} from the corrected asteroseismic scaling relation for the Kepler target stars \protect\citep[APKASC-2,][]{2018ApJS..239...32P}. The effective temperatures and metallicities of these stars are from the APOKASC-2 catalogue, and the corrected distances are from Gaia DR3 \protect\citep{Gaia2021}. Squares indicate the oscillating components of the EBs, while circles and filled circles mark the sides of the triangle in the MZ diagram. Most of these stars are located inside the triangle. {In the lower panel, the RGBs with the $\alpha$-elements enhancement ([$\alpha$/M]$>$ 0.1) are plotted (filled circle with dark grey). The left corner of the triangle is occupied by these RGBs. Six of the SLO RGs are in this region.}}
    \label{fig:MZ1}
\end{figure}

\subsection{Asteroseismic properties of the SLO components}
The oscillation frequencies of the SLO components
are derived from \textit{Kepler} photometric data \citep{2011AJ....142..112B}. The derived values of $\nu_{\rm max}$ and ${\Delta \nu}$ from the observations are compiled from the literature.  These parameters of the SLO components are listed in Table \ref{tab:obs}. For KIC 8410637 (Tek Ayak\footnote{It is the name of the crow in the Ege University campus that has the ability to live despite having only one leg.}) 
and KIC 9540226, the frequencies of the individual oscillation modes with $l=0$, $1$, and $2$ are taken from \cite{2018MNRAS.478.4669T}.


\begin{table}
  \centering
  \caption{ Observational asteroseismic properties of the SLO RGs in the EBs.  Ref. No.: 1- \protect\cite{2022A&A...668A..82B}, 2- \protect\cite{2016ApJ...832..121G}, 5- \protect\cite{2018MNRAS.478.4669T} and 7- \protect\cite{2010ApJ...713L.187H}. The columns are shown with KIC ID, large separation (\Dnu), frequency of maximum amplitude (\numax), and reference (Ref).}
    \begin{tabular}{lccc}
\hline
  KIC    &  \Dnu       & \numax        & Ref  \\
         &  $\mu$Hz    &  $\mu$Hz      &     \\
 \hline
 4054905& 5.40$\pm$0.15& 48.44$\pm$0.55   &  1  \\
 4663623& 5.21$\pm$0.02& 54.09$\pm$0.24   &  2  \\
 5786154& 3.52$\pm$0.01& 29.75$\pm$0.16   &  2  \\
 7037405& 2.79$\pm$0.01& 21.75$\pm$0.14   &  2  \\
 7377422& 4.64$\pm$0.05& 40.10$\pm$2.10   &  2  \\
 8410637& 4.64$\pm$0.017& 46.00$\pm$0.19   &  2  \\
    "   & 4.564$\pm$0.004& 46.4 $\pm$0.3    &  5  \\
    "   & 4.5$\pm$0.1& 45.2 $\pm$1.3        &  7  \\ 
 8430105& 7.14$\pm$0.03& 76.70$\pm$0.57   &  2  \\
 9246715& 8.31$\pm$0.02&106.40$\pm$0.80   &  2  \\
 9540226& 3.22$\pm$0.01 &27.07$\pm$0.15   &  2  \\
    "   & 3.19$\pm$0.01 &26.7 $\pm$0.2    &  5  \\
 9970396& 6.32$\pm$0.01 &63.70$\pm$0.16   &  2  \\
10001167& 2.76$\pm$0.01 &19.90$\pm$0.09   &  2  \\
\hline
    \end{tabular}%
  \label{tab:obs}%
\end{table}%

\subsection{The total metallicity of 11 SLO components and their position in the MZ diagram}
\label{sec:11SLOMZ}
{
The [Fe/H] values obtained from the spectroscopic data are given in Table \ref{tab:EBobs}. The surface metallicities ($Z_{\rm s}$) of the RG components are calculated from these [Fe/H] values:
\begin{equation}
\label{eq:ZsFeH}
Z_{\rm s}=10^{\rm [Fe/H]} Z_{\sun}, 
\end{equation}
where $Z_{\sun}$ is the solar metallicity and taken as 0.0134 \citep{Asplund2009}. The $Z_{\rm s}$ of the systems varies between 0.0025 and 0.0194. The initial metallicity ($Z_0$) for RGs is very close to $Z_{\rm s}$ (Method I, see Section \ref{sec:MethodI}) as they undergo the first dredge-up.}

{Three basic parameters related to the structure of stars are obtained from their spectral data. These are $\teff$, {\rm [Fe/H]}, and $\log (g)$ values. Different spectral studies on the same stars often give different results for these parameters. One of the main problems is which of these will be used in model studies. Perhaps the most appropriate thing is to look for solutions for each. In this case, we have to spend more effort and more time. A better approach is to test whether there is a relationship for these three parameters in the given results and reach a conclusion from this analysis (see Appendiz \ref{sec:kic5786154}).}

The MZ diagram plays a crucial role in determining the structural and evolutionary status of RG stars. The MZ diagram constructed by \cite{2023MNRAS.518.5552Y} with the mass calculated from the corrected asteroseismic scaling relations for RGs is shown in Fig. \ref{fig:MZ1}. 
In the MZ diagram, four of the 11 binary systems have oscillating components (KIC 4054905, KIC 8430105, KIC 9540226 and KIC 10001167) located near the base boundary of the triangular structure. The oscillating component of KIC 9246715 is located in the region containing few stars with large masses. The oscillating components of the remaining six systems are located inside the triangle. 

{ This diagram also provides insight into the ages of the stars. 
The left boundary of the triangle is mainly defined by stars that have only recently reached the RG phase. In the Galactic disk, metal-rich stars are generally younger because chemical enrichment requires time. At the same time, stars also require sufficient evolutionary time to reach the RG branch. The combined effect of these two constraints produces the prominent left boundary of the triangle \citep{2023MNRAS.518.5552Y}. 

In contrast, stars near the lower boundary are the oldest systems and can be interpreted as stars approaching the end of the RG phase. Stars located below the lower boundary may have experienced mass gain during their previous evolution, possibly through the engulfment of companions. 
The KIC 4054905 and KIC 10001167 systems, whose SLO components have the lowest $M$ and $Z$ values, are therefore among the oldest systems in the sample.}

{Kepler target stars include both thin and thick disk members. Besides their kinematic properties \citep{2022ApJ...932...28V}, one of the primary distinguishing characteristics of stars belonging to the two disks is their abundances of $\alpha$-group elements ([$\alpha$/Fe] or [$\alpha$/M]) relative to Fe (or metallicity = M). Thick disk member stars have an abundance of [$\alpha$/M] greater than 0.1 dex \citep{2021A&A...645A..85M}. In the lower panel of Fig. {\ref{fig:MZ1}}, the background stars are marked according to their [$\alpha$/M]. The [$\alpha$/M] values are taken from the APOGEE DR17 data in the APOKASC-3 catalogue \citep{2022ApJS..259...35A,2025ApJS..276...69P}. Stars with [$\alpha$/M] > 0.1 occupy the lower left corner of the triangle in the MZ diagram. Since these are the lowest-mass stars within the triangle, they are naturally also the oldest. We note that almost all of the stars below the triangle, which may be candidate stars that gained mass in the past, are rich in alpha elements.  
Six of the SLO RGs in EBs are in the $\alpha$-rich region of the triangle.  
}
\section{Model descriptions and modelling of the EB components}
\label{sec:sec3}

{ The present analysis is not based on interpolation within precomputed stellar model grids. Instead, stellar evolution models are constructed individually for each binary component using the observed masses and trial chemical compositions. For every assumed chemical composition, evolutionary tracks are computed with {\small MESA} until the observed luminosities are reproduced, yielding the corresponding stellar ages.

The primary objective of this modelling procedure is to determine the ages $t$ of these systems. The resulting models also constrain the initial chemical compositions and provide insights into possible mass transfer events. Furthermore, comparisons between the asteroseismic properties of the models and the observations offer an additional diagnostic for assessing the reliability of the stellar structure models and the asteroseismic scaling relations.

It is generally assumed that the components of a binary system are coeval and share the same initial chemical composition. Under this assumption, the stellar age is determined by requiring that both components are reproduced at a common evolutionary state within a consistent stellar modelling framework.}

All stellar models are constructed using the {\small MESA} code \citep{Paxton2011, Paxton2013, Paxton2015, Paxton2018, Paxton2019, Jermyn2023} (r23.05.1). Because the target stars rotate very slowly, non-rotating models are adopted. Convection is treated using the standard mixing-length theory of \citet{Bohm1958}. High-temperature opacities ($T>10^4$ K) are taken from OPAL tables \citep{Iglesias1993, Iglesias1996}, while low-temperature opacities are adopted from \citet{Ferguson2005}. Element diffusion is included following \citet{1986ApJS...61..177P}. Pre-main-sequence evolution is included in all model calculations.

Atmospheric boundary conditions are computed using the MESA options \texttt{atm\_option = 'table'} and \texttt{atm\_table = 'photosphere'}. In this setup, $T_{\rm surf}$ and $P_{\rm surf}$ are obtained by interpolation from atmospheric tables constructed using PHOENIX-based models \citep{1999ApJ...512..377H,1999ApJ...525..871H} and \cite{2003IAUS..210P.A20C}.

{ In the Hertzsprung--Russell diagram, the CHeB phase occupies a relatively narrow and degenerate region, where multiple model solutions may exist for a given mass and chemical composition. For such cases, we adopt a forward-modelling strategy in which the system age is anchored primarily by the red-giant component, and the secondary component is used to refine the solution under the coevality constraint.

\subsection{Method I}
\label{sec:MethodI}

In Method I, we adopt the observed metallicity $Z_0$ (i.e. $Z_{\rm s}$; Table \ref{tab:EBobs}) as an initial constraint. For a given choice of the initial helium abundance $Y_0$, stellar evolution models are computed for both components using MESA. The stellar age $t$ is then determined by requiring that both components simultaneously reproduce the observed luminosities at a common evolutionary state.

We initially assume $\alpha = \alpha_\odot$. The mixing-length parameters of the primary ($\alpha_{\rm A}$) and secondary ($\alpha_{\rm B}$) components are subsequently refined by matching the observed radii and effective temperatures. In this framework, $Y_0$ primarily controls the coevality condition through the luminosity constraint, while $\alpha$ acts as a secondary structural parameter affecting the radius--temperature relation.

The observational constraints and model parameters used in the modelling procedure are listed in Table \ref{tab:methods_parameters}. Excluding the asteroseismic quantities, the system is constrained by a set of observables that is comparable in number to the model degrees of freedom, allowing the stellar parameters to be determined without fixing $Y_0$ or the mixing-length parameters to their solar values. In this framework, the stellar age $t$ is determined from the requirement of simultaneous reproduction of the observed luminosities of both components at a common evolutionary state, while $Y_0$ and $\alpha$ are subsequently constrained through the coevality and radius--temperature conditions, respectively. The consistency of the resulting solution therefore depends on the adequacy of the adopted stellar physics in MESA. In this sense, agreement between the model predictions and the observed asteroseismic quantities provides an additional and independent validation of the inferred stellar structure.

{Although Method I allows solutions with very low $Y_0$ values, no lower limit is imposed during the calibration procedure. In some systems, the resulting $Y_0$ becomes smaller than the primordial helium abundance ($Y_{\rm p}$; see Table \ref{tab:Yol12_t9}). While such models formally satisfy the observational constraints, they are unlikely to represent physically meaningful solutions because Galactic chemical evolution predicts that helium abundance should increase together with metallicity. These cases probably reflect limitations in the adopted input physics and/or uncertainties in the observational parameters.}

\subsection{Method II}

In Method II, we determine the stellar ages by relaxing the direct use of observed metallicities and instead introducing a physically motivated relation between the initial helium and metal abundances. We assume that chemical enrichment in the Galactic disc follows
$Y_0 = Y_{\rm p} + c_{YZ} Z_0,$
where $Y_{\rm p} = 0.2471$ is the primordial helium abundance \citep{2020A&A...641A...6P} and $c_{YZ}$ is a parameter to be constrained. For each assumed value of $c_{YZ}$, evolutionary models are computed for both components and the stellar ages are determined by enforcing the coevality condition ($t_{\rm A} = t_{\rm B}$) under simultaneous luminosity constraints. This procedure is repeated for $c_{YZ} = 1.5, 2,$ and $3$, allowing us to assess the sensitivity of the inferred ages to the adopted chemical enrichment relation. The value of $c_{YZ}$ that yields the most consistent solutions across the sample is found to be $c_{YZ} \simeq 2$.

Although a value of $c_{YZ} \approx 2$ is widely adopted (e.g. \cite{2021MNRAS.501..383T} for the Hyades), values in the range $1 \lesssim c_{YZ} \lesssim 5$, and in some cases even higher values for low-metallicity populations, have been reported in the literature \citep{2007MNRAS.382.1516C}. This reflects the still open question of whether a single universal relation between $Y_0$ and $Z_0$ can adequately describe Galactic chemical evolution \citep{2019A&A...630A.125V}.

{\small\addtolength{\tabcolsep}{-3pt}
\begin{table}
 \centering
\caption{The parameters fixed as the observed values and the free parameters used in Method I and Method II.}
\begin{tabular}{lll}
\hline
{}  & {Method I} & {Method II} \\
\hline
\text{Observational constr. } & $M_{\rm A}$, $M_{\rm B}$, $Z_0$ & $M_{\rm A}$, $M_{\rm B}$, ($\Delta Y=2\Delta Z$) \\
& $T_{\rm effA}$, $T_{\rm effB}$, $L_{\rm A}$, $L_{\rm B}$ & $T_{\rm effA}$, $T_{\rm effB}$, $L_{\rm A}$, $L_{\rm B}$\\
\hline
\text{Model fixed param.} & $M_{\rm A}$, $M_{\rm B}$, $Z_0$, $t_{\rm A}=t_{\rm B}$  & $M_{\rm A}$, $M_{\rm B}$, $Y_0(Z_0)$, $t_{\rm A}=t_{\rm B}$  \\
\text{Model free param.} & $\alpha_{\rm A}$, $\alpha_{\rm B}$, $Y_0$ & $\alpha_{\rm A}$, $\alpha_{\rm B}$, $Z_0$  \\
\hline
\end{tabular}%
\label{tab:methods_parameters}    
\end{table}}%

In Method I, all relevant structural parameters, including $Y_0$ and the mixing-length parameters, are constrained within the modelling procedure. In Method II, the relation between $Y_0$ and $Z_0$ is prescribed through $c_{YZ}$, which reduces the number of independent free parameters by one (see Table \ref{tab:methods_parameters}). The comparison of the two approaches therefore provides constraints on the range of initial chemical compositions that are consistent with both stellar evolution and observational data.
}
 
\section{Results and discussion}
\label{sec:sec4}

\begin{figure*}
    \centering
\includegraphics[width=\textwidth]{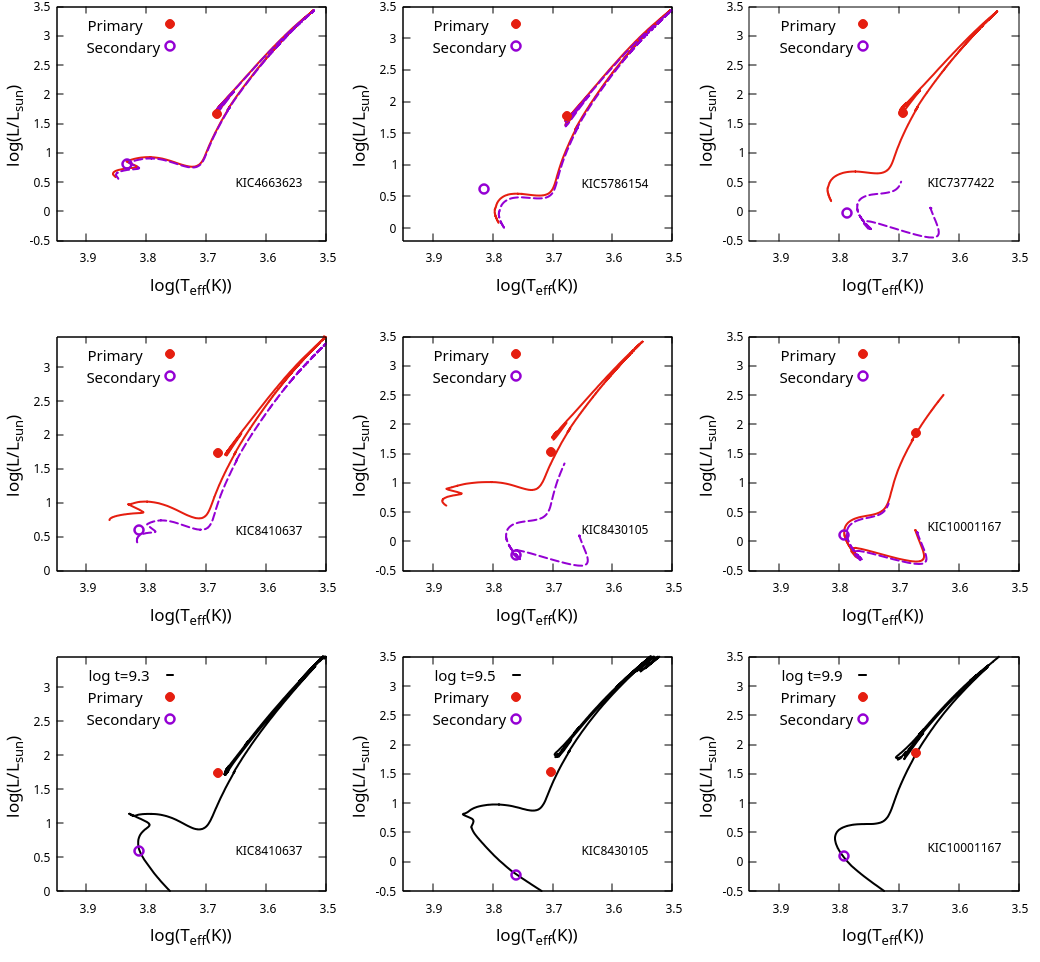}
    \caption{ HRDs of the components of the SLO EBs. The filled circles represent the primary components and the open circles represent the secondary components. The evolutionary tracks are obtained from stellar evolution models computed for masses closest to the observed values of $M_{\rm A}$ and $M_{\rm B}$ and for the adopted metallicity $Z$. 
      Two different observational data sets are shown for some systems. Since the solar value is adopted for the mixing-length parameter in these illustrative tracks, temperature offsets between the evolutionary tracks and the observational positions may occur. 
      In terms of luminosity, the secondary components of five systems are located around the MS, whereas the secondary component of KIC 5786154 is more luminous than expected for the adopted metallicity, suggesting that the actual $Z$ value of this system may be lower than the observed one. { For the low-mass secondary components of KIC 7377422, KIC 8430105, and KIC 10001167, the evolutionary tracks also include the pre-main-sequence phase.} In the lower panel, the isochrones for KIC 8410637, KIC 8430105, and KIC 10001167 are plotted. Although the isochrones appear to reproduce the positions of both components well, the stellar masses at the fitted locations differ significantly from the observed dynamical masses (see text).}
    \label{fig:HRD}
\end{figure*}

We construct interior models for the components of the double-lined binaries in the sample of 11 EBs. For several systems, more than one observational solution from dynamical analyses is available. All compiled observational values are listed in Table \ref{tab:EBobs}. In general, the different observational studies are mutually consistent, although significant differences exist between some of the reported $T_{\rm eff}$ values. The bold reference numbers in the last column of Table \ref{tab:EBobs} indicate the data sets adopted in the present modelling.

Preference is given to the observational sets with smaller uncertainties in the component masses and radii. If no satisfactory solution is obtained with Method I for a given set, alternative observational sets are tested. For four systems (KIC 8410637, KIC 8430105, KIC 9540226, and KIC 10001167), interior models are constructed for two independent observational data sets.

Most of the SLO stars in the EBs are located near the base of the triangle in the MZ diagram. Two systems (KIC 5786154 and KIC 8410637) lie significantly above the baseline, while four systems (KIC 4054905, KIC 8430105, KIC 9540226, and KIC 10001167) are located outside or close to the border of the triangle. Therefore, except perhaps for KIC 4054905, the SLO RGs in these binaries can be interpreted as evolving approximately with constant mass.

For all binaries, we first plot HRDs using evolutionary tracks computed for selected $Z$ values and the solar mixing-length parameter. In these illustrative model sequences, the helium abundance is calculated from the chemical enrichment relation with $c_{YZ}=2$:
    $Y = Y_{\rm p} + 2Z$.

The HRDs of six binaries are shown in the upper and middle panels of Fig. \ref{fig:HRD}. For four binaries (KIC 4663623, KIC 8410637, KIC 8430105, and KIC 10001167), the evolutionary tracks pass through or very close to the observed positions of both components. However, this agreement alone does not guarantee a simultaneous solution for the system. The important point is whether the component models reach the observed positions at the same age. 

For two systems, KIC 5786154 and KIC 7377422, the agreement between the tracks and the observed positions is poorer. In particular, the model luminosity of the secondary component of KIC 5786154 is slightly higher than observed. Such discrepancies can be reduced either by increasing $Y$ for a fixed $Z$ or by decreasing $Z$ for a given $Y-Z$ relation. For KIC 7377422 B, the mixing-length parameter $\alpha$ also plays an important role in reproducing the observed position in the HRD.

In the lower panel of Fig. \ref{fig:HRD}, the isochrones for KIC 8410637, KIC 8430105, and KIC 10001167 are plotted. At first sight, the isochrones \citep{2016ApJ...823..102C,2016ApJS..222....8D} appear to reproduce the observed positions of both components very well. However, projecting a multidimensional problem on to a two-dimensional diagram can easily produce misleading visual agreement because one parameter remains hidden. In the upper and middle panels, the hidden parameter is the stellar age, whereas in the lower panel it is the stellar mass.

For example, although the isochrone for KIC 10001167 reproduces the observed positions of both components remarkably well, the corresponding isochrone masses ($M_{\rm A}=0.96$ and $M_{\rm B}=0.85~\MS$) differ significantly from the dynamical masses reported by \cite{2016ApJ...832..121G} ($M_{\rm A}=0.81$ and $M_{\rm B}=0.79~\MS$). On the other hand, the isochrone masses are in much better agreement with the revised masses reported by \cite{2025A&A...699A.152T} ($M_{\rm A}=0.9337$ and $M_{\rm B}=0.8388~\MS$).

The component masses of KIC 4054905, KIC 4663623, and KIC 9246715 are so similar that Methods I and II cannot be applied reliably to these systems. For these binaries, solutions are obtained only from models of the primary components. The corresponding ages are found to be 9.33, 3.28, and 0.78 Gyr for KIC 4054905, KIC 4663623 (see also Section \ref{sec:4663623}), and KIC 9246715, respectively.

Four of the six SLO stars shown in Fig. \ref{fig:HRD} are located inside the triangular region of the MZ diagram (see Fig. \ref{fig:MZ1}), whereas KIC 8430105 and KIC 10001167 are located close to the border of this region.

\subsection{Method I: determination of $Y_0$ from the coevality condition --- the case of KIC 7377422}
\label{sec:MethodIResults}

For the application of Method I, we initially adopt $\alpha_{\rm A}=\alpha_{\rm B}=\alpha_{\sun}$ because the mixing-length parameter has only a minor effect on the luminosity of a stellar model at a given evolutionary stage. The initial metallicity is taken as the observed surface metallicity, i.e. $Z_0=Z_{\rm s}$ (Table \ref{tab:EBobs}). 

{ For a trial value of $Y_0$, stellar evolution models are constructed separately for the primary and secondary components using their observed masses. The ages of the component models are then determined from the condition that the model luminosities reproduce the observed luminosities. Matching the positions of the two stars in the HRD is therefore necessary but not sufficient. The additional requirement for a physically acceptable solution is the coevality condition
    $t_{\rm A}=t_{\rm B}$.
}

\begin{figure}
    \centering
\includegraphics[width=1.15\linewidth]{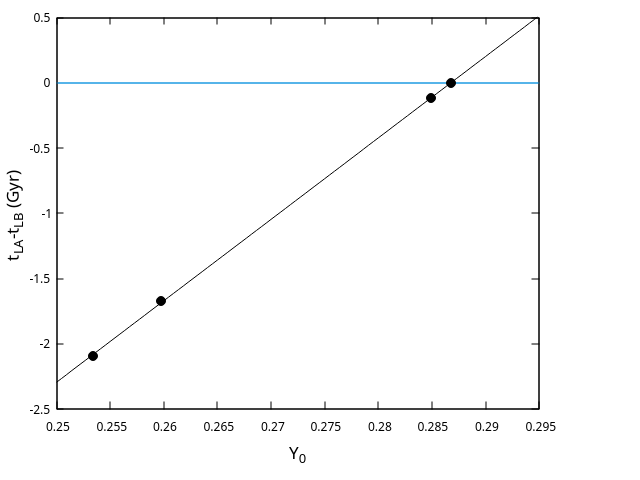}    
\includegraphics[width=1.15\linewidth]{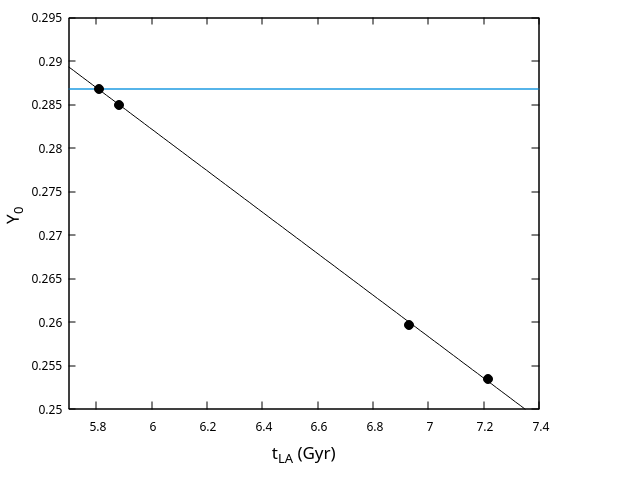}
    \caption{{ Application of Method I to KIC 7377422.} 
    (a) The age difference between the primary and secondary component models is plotted against $Y_0$. A nearly linear relation is obtained, and the coevality condition ($t_{\rm A}=t_{\rm B}$) is satisfied at $Y_0=0.2868$. 
    (b) The corresponding system age as a function of $Y_0$. For the coeval solution, the system age is 5.79 Gyr.}
    \label{fig:y0t9}
\end{figure}

Both component ages depend on $Y_0$, although the age of the secondary component is generally more sensitive to changes in $Y_0$ than that of the primary. Figure \ref{fig:y0t9}(a) shows the variation of $t_{\rm A}-t_{\rm B}$ with $Y_0$ for KIC 7377422. A nearly linear relation is obtained between the age difference and $Y_0$. The coevality condition is satisfied at $Y_0=0.2868$, where $t_{\rm A}-t_{\rm B}=0$. The corresponding system age is 5.79 Gyr, as shown in Fig. \ref{fig:y0t9}(b) and listed in Table \ref{tab:Yol12_t9}.

{ After determining the age, the mixing-length parameters are adjusted to reduce the differences between the observed and model effective temperatures. Small changes in the ages produced during this step are compensated by minor adjustments in $Y_0$. For KIC 7377422, the final calibrated values are} $\alpha_{\rm A}=2.1731$ and $\alpha_{\rm B}=2.1401$.

Method I is also applied to the other EBs in the sample. The results are summarized in Table \ref{tab:Yol12_t9}. No satisfactory solution is obtained for KIC 4054905, KIC 4663623, and KIC 9246715 because of the very similar masses of their components. In addition, the two observational solutions tested for KIC 10001167 yield unrealistically low helium abundances ($Y_0<0.22<Y_{\rm p}$). 

For KIC 5786154, the second solution with $Z_0=0.0057$ appears more physically plausible than the first solution, particularly in view of the corresponding Method II result.

Method I yields $Y_0$ values in the range 0.2381--0.3039. Among these systems, only KIC 7037405 gives a helium abundance below the primordial value $Y_{\rm p}$. The youngest system in the sample is KIC 8410637 (Tek Ayak) with an age of 2.17 Gyr, whereas the oldest system is KIC 7377422 with an age of 5.79 Gyr. The calibrated mixing-length parameters span the ranges $\alpha_{\rm A}=1.7811$--2.2358 and $\alpha_{\rm B}=1.4311$--2.5666.

We also compute ages ($t_{\rm f}$) for the SLO components using the fitting formula derived from the MESA models (Appendix \ref{sec:AppA}). { The fitting formula gives the stellar age as a function of $M$, $Z_0$, and $R$, and is based on stellar models computed with the chemical enrichment relation $\Delta Y=2\Delta Z$. In general, the agreement between $t_{\rm f}$ and the ages obtained from Method I improves when the inferred chemical composition of the system is close to the relation corresponding to} $c_{YZ}\approx2$.  

\subsection{Method II: determination of $Z_0$ from the coevality condition --- the case of KIC 7377422}

Using Method II, we calibrated the component models of eight { double-lined spectroscopic binary (SB2) systems with sufficiently precise observational parameters (Table \ref{tab:EBobs}). In this method, the initial helium abundance is linked to metallicity through the chemical enrichment relation
$Y_0 = Y_{\rm p} + 2 Z_0 ,$
and the coevality condition is used to determine the chemical composition and age of the system simultaneously.} The final model parameters obtained for all systems are listed in Table \ref{tab:Yol12_t9}.

To illustrate the method, we again consider the KIC 7377422 system. In the HRD shown in Fig. \ref{fig:HRD}, the evolutionary tracks were computed using $Z=0.0075$, which is the closest grid value to the observed metallicity ($Z \simeq 0.0068$). For this metallicity, the ages of the two components differ significantly:
$t_{\rm A}-t_{\rm B}=-2.475~{\rm Gyr}$
when the primary component is assumed to be on the RGB, and
$t_{\rm A}-t_{\rm B}=-2.407~{\rm Gyr}$
when it is assumed to be in the RC phase. Therefore, for the adopted $Y_0-Z_0$ relation, the metallicity must be reduced to satisfy the coevality condition.

\begin{figure}
    \centering
\includegraphics[width=1.15\linewidth]{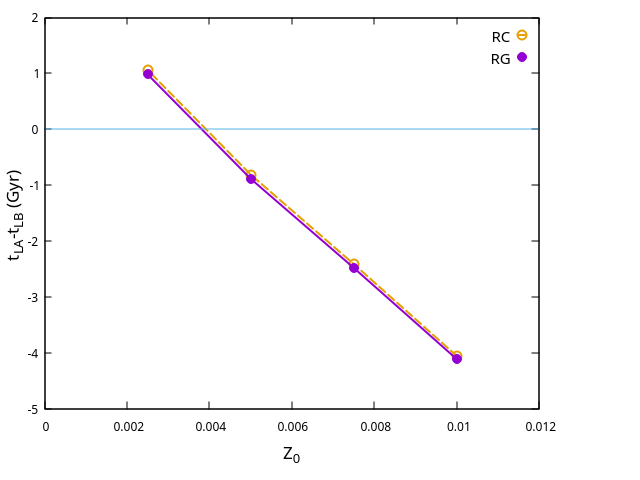}
\includegraphics[width=1.15\linewidth]{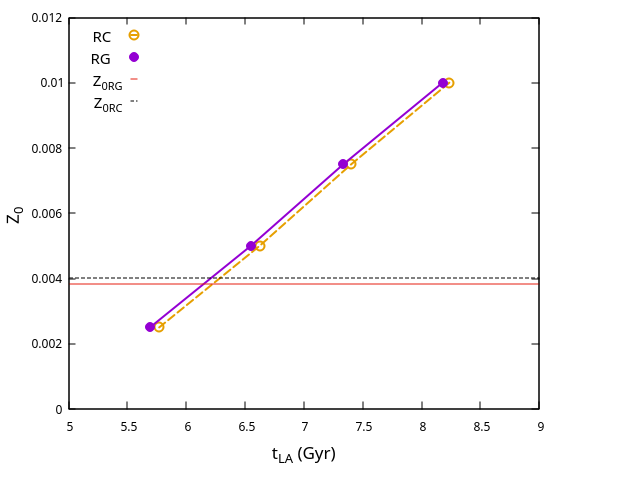}
    \caption{
    (a) Age difference between the components of KIC 7377422 plotted against $Z_0$ for Method II. Circles and filled circles represent the RGB and RC interpretations of the primary component, respectively. The coevality condition is satisfied near $Z_0 \simeq 0.0037$.
    (b) Corresponding system age as a function of $Z_0$. The final iterative solution yields an age of approximately 6.13 Gyr.
    }
    \label{fig:z0t9}
\end{figure}

The variation of the age difference with $Z_0$ is shown in the a) panel of Fig. \ref{fig:z0t9}. As $Z_0$ decreases, the signed age difference $(t_{\rm A}-t_{\rm B})$ changes nearly linearly and becomes zero at the coeval solution. For the RGB interpretation of the primary component, the equality $t_{\rm A}=t_{\rm B}$ is obtained at ${Z_0=0.00382}$. For the RC interpretation, the corresponding solution is ${Z_0=0.00402}$. These two values are very close because the transition from RGB to RC occurs rapidly in stellar evolution.

The b) panel of Fig. \ref{fig:z0t9} shows the dependence of the system age on $Z_0$. The corresponding ages are 6.145 Gyr for the RGB case and 6.290 Gyr for the RC case. After obtaining these luminosity-based solutions, we refined the models by adjusting the mixing-length parameters to reproduce the observed effective temperatures. Since changes in $\alpha$ slightly modify the evolutionary ages, the $Z_0$ values were recalculated iteratively. The final solution for KIC 7377422 is
${Z_0 = 0.00368,\quad
t = 6.13~{\rm Gyr},}$
with
$\alpha_{\rm A}=2.0129,\quad
\alpha_{\rm B}=1.9412.$

Method II yields $Y_0$ values in the range 0.2545--0.2797, which is significantly narrower than the range obtained with Method I. All derived $Y_0$ values are greater than the primordial helium abundance $Y_{\rm p}$. The youngest system is again KIC 8410637 (Tek Ayak), with an age of 2.30 Gyr, whereas the oldest accepted solution corresponds to KIC 10001167, with an age of 10.08 Gyr.
\begin{table*}
\small\addtolength{\tabcolsep}{-2pt}
    \centering
    \caption{Solutions obtained with Methods I and II for the eight EBs from the interior models of the component stars. The model luminosities are only equal to the observational ones for a single $Y_0$ (Method I) or $Z_0$ (Method II) value. The second and third columns list $Z_{0}$ and $Y_0$, respectively. The primary components are assumed to be RG. The corresponding age ($t$) is listed in the fourth column. {$t_{\rm f}$, the fitting formula for the age derived from the RG models (see Appendix), is given in the fifth column. The  $\alpha_{\rm A}$ and  $\alpha_{\rm B}$ values obtained from the fit of the model  $T_{\rm eff}$ to the observational  $T_{\rm eff}$s are given in the sixth and seventh columns. $c_{YZ}$, the factor for the chemical enrichment law between $Z$ and $Y$, is given in the eighth column: $Y_{0}=Y_{\rm p}+c_{YZ}Z_0$. Columns 9$-$12 list modelled and observed values of \numax~ and \Dnu~}.
}
\begin{tabular}{rccccccccccc}
\hline
KIC ID & $Z_{\rm 0}$ & $Y_{\rm 0}$ & $t_{\rm 9}$ & $t_{\rm f}$   & $\alpha_{\rm A}$ & $\alpha_{\rm B}$ & $c_{YZ}$ &  \numax (model)    &  \Dnu(model)       & \numax        &  \Dnu   \\
&             &             &        Gyr      &        Gyr      &                 &                   &  
         &$\mu$Hz    &  $\mu$Hz &  $\mu$Hz    &  $\mu$Hz\\
\hline
Method I &  & & & & & & & & & \\
        \hline
5786154  & 0.0117 $\pm$ 0.0015 & 0.3484 $\pm$ 0.0041 & 4.83$^{+1.08}_{-0.91}$ & 8.22 &2.1451 & 2.9911 & 8.66  & 29.46$^{+1.44}_{-1.45}$ & 3.549$^{+0.094}_{-0.093}$ & 29.75$\pm$0.16 & 3.523$\pm$0.014    \\
   "     & 0.0057 $\pm$ 0.0015 & 0.2803 $\pm$ 0.0041 & 5.67$^{+1.24}_{-1.08}$ & 6.73 & 1.9544 & 2.5666 & 5.83  & 28.41$^{+1.39}_{-1.40}$ & 3.516$^{+0.092}_{-0.093}$ & 29.75$\pm$0.16 & 3.523$\pm$0.014   \\ 
7037405  & 0.0072 $\pm$ 0.0015 & 0.2381 $\pm$ 0.0041 & 5.75$^{+0.52}_{-0.48}$ & 4.86 &1.7811 & 1.4311 & -1.30 & 20.74$^{+0.37}_{-0.37}$ & 2.692$^{+0.026}_{-0.026}$ & 21.75$\pm$0.14 & 2.792$\pm$0.012    \\
7377422  & 0.0063 $\pm$ 0.0008 & 0.2868 $\pm$ 0.0022 & 5.79$^{+1.76}_{-1.44}$ & 7.13 & 2.1731 & 2.1401 & 6.30  & 39.74$^{+2.61}_{-2.56}$ & 4.604$^{+0.155}_{-0.153}$ & 40.10$\pm$2.10 & 4.643$\pm$0.052    \\
8410637  & 0.0194 $\pm$ 0.0013 & 0.3039 $\pm$ 0.0035 & 2.17$^{+0.15}_{-0.15}$ & 2.40 & 2.2358 & 2.0696 & 2.93  & 48.00$^{+1.01}_{-0.99}$ & 4.673$^{+0.061}_{-0.060}$ & 46.00$\pm$0.19 & 4.641$\pm$0.017    \\
{ 8410637T}  & 0.0140 $\pm$ 0.0013 & 0.1933  & 4.54 & 2.69 & --- & --- & -3.84  & --- & --- & 46.00$\pm$0.19 & 4.641$\pm$0.017    \\
 { 8410637TzT}  & 0.0194 $\pm$ 0.0013 & 0.2868 $\pm$ 0.0035 & 2.99$^{+0.15}_{-0.15}$ & 2.69 & 1.9680 & 2.2239 & 2.05  & 46.80$^{+1.01}_{-0.99}$ & 4.649$^{+0.061}_{-0.060}$ & 46.00$\pm$0.19 & 4.641$\pm$0.017    \\   
8430105  & 0.0043 $\pm$ 0.0004 & 0.2482 $\pm$ 0.0011 & 3.08$^{+0.19}_{-0.18}$ & 2.91  &1.9311 & 1.8123 & 0.25  & 74.12$^{+1.12}_{-1.11}$ & 7.103$^{+0.065}_{-0.064}$ & 76.70$\pm$0.57 & 7.138$\pm$0.031    \\
{ 8430105T}  & 0.0046 $\pm$ 0.0012 & 0.2451  & 3.74 & 3.46 & --- & --- & -0.43  & --- & --- & 76.70$\pm$0.57 & 7.138$\pm$0.031    \\
9540226  & 0.0063 $\pm$ 0.0006 & 0.2740 $\pm$ 0.0016 & 2.81$^{+0.39}_{-0.35}$ & 3.10 & 1.8311 & 2.0102 & 4.28  & 28.26$^{+0.89}_{-0.86}$ & 3.353$^{+0.052}_{-0.052}$ & 27.07$\pm$0.15 & 3.216$\pm$0.013    \\
{ 9540226B}  & 0.0079 $\pm$ 0.0016 & 0.26045 $\pm$ 0.0016 & 2.94$^{+0.36}_{-0.32}$ & 2.92 & 1.9145 & 1.6974 & 1.69  & 28.05$^{+0.36}_{-0.31}$ & 3.307$^{+0.055}_{-0.055}$ & 27.07$\pm$0.15 & 3.216$\pm$0.013    \\
9970396  & 0.0085 $\pm$ 0.0004 & 0.2633 $\pm$ 0.0011 & 5.04$^{+0.23}_{-0.23}$ & 5.10 & 2.0218 & 1.8935 & 1.91  & 62.12$^{+1.12}_{-1.11}$ & 6.236$^{+0.077}_{-0.075}$ & 63.70$\pm$0.16 & 6.320$\pm$0.010    \\
10001167 & 0.0027 $\pm$ 0.0002 & 0.1518   & ---  & 14.22 & ---    & ---    & -35.3 & ---   & ---   & 19.90$\pm$ 0.09 & 2.762$\pm$ 0.012   \\
{ 10001167T} & 0.0025 $\pm$ 0.0006 & 0.2169   & ---  & 8.41 & ---    & ---    & -12.1 & ---   & ---   & 19.90$\pm$ 0.09 & 2.762$\pm$ 0.012   \\
        \hline
Method II &  & & & & & & & & & \\
\hline
5786154  & 0.0039 $\pm$ 0.0015 & 0.2549 $\pm$ 0.0030 & 6.04$^{+1.50}_{-1.31}$ & 6.06 & 1.8311 & 2.2940 & 2.00 & 28.01$^{+1.37}_{-1.38}$ &  3.511$^{+0.092}_{-0.093}$ & 29.75$\pm$0.16 & 3.523$\pm$0.014 \\
7037405  & 0.0093 $\pm$ 0.0015 & 0.2657 $\pm$ 0.0030 & 5.42$^{+0.41}_{-0.40}$ & 5.46 & 1.8619 & 1.5848 & 2.00 & 21.22$^{+0.38}_{-0.37}$ &  2.703$^{+0.026}_{-0.026}$ & 21.75$\pm$0.14 & 2.792$\pm$0.012 \\
7377422  & 0.0037 $\pm$ 0.0008 & 0.2545 $\pm$ 0.0016 & 6.13$^{+1.85}_{-1.56}$ & 6.15 & 2.0129 & 1.9412 & 2.00 & 39.14$^{+2.57}_{-2.52}$ &  4.629$^{+0.156}_{-0.154}$ & 40.10$\pm$2.10 & 4.643$\pm$0.052 \\
8410637  & 0.0163 $\pm$ 0.0013 & 0.2797 $\pm$ 0.0026 & 2.30$^{+0.17}_{-0.16}$ & 2.29 & 2.1884 & 1.9175 & 2.00 & 47.47$^{+1.00}_{-0.98}$ &  4.695$^{+0.062}_{-0.060}$ & 46.00$\pm$0.19 & 4.641$\pm$0.017 \\
{ 8410637T}  & 0.0326 $\pm$ 0.0013 & 0.3122 $\pm$ 0.0026 & 3.29 & 3.41 & --- & --- & 2.00 & --- &  --- & 46.00$\pm$0.19 & 4.641$\pm$0.017 \\
 { 8410637TzT} & 0.0192 $\pm$ 0.0013 & 0.2855 $\pm$ 0.0035 & 2.99$^{+0.15}_{-0.15}$ & 2.69 & 1.9680 & 2.2239 & 2.00  & 46.82$^{+1.01}_{-0.99}$ & 4.654$^{+0.061}_{-0.060}$ & 46.00$\pm$0.19 & 4.641$\pm$0.017    \\   
8430105 & 0.0051 $\pm$ 0.0004 & 0.2572 $\pm$ 0.0008 & 3.03$^{+0.18}_{-0.17}$ & 3.05 & 1.9908 & 1.8599 & 2.00 & 74.70$^{+1.13}_{-1.12}$ &  7.110$^{+0.065}_{-0.064}$ & 76.70$\pm$0.57 & 7.138$\pm$0.031 \\
{ 8430105T}  & 0.0058 $\pm$ 0.0012 & 0.2586 $\pm$ 0.0008 & 3.66$^{+0.24}_{-0.27}$ & 3.69 & 1.9849 & 1.9352 & 2.00 & 75.47$^{+0.99}_{-0.98}$ &  7.221$^{+0.048}_{-0.047}$ & 76.70$\pm$0.57 & 7.138$\pm$0.031 \\
9540226  & 0.0048 $\pm$ 0.0006 & 0.2566 $\pm$ 0.0012 & 2.88$^{+0.41}_{-0.38}$ & 2.88 & 1.9073 & 1.7540 & 2.00 & 28.08$^{+0.89}_{-0.86}$ &  3.366$^{+0.052}_{-0.053}$ & 27.07$\pm$0.15 & 3.216$\pm$0.013 \\
{ 9540226B}  & 0.0082 $\pm$ 0.0016 & 0.26344 $\pm$ 0.0012 & 2.92$^{+0.33}_{-0.32}$ & 2.95 &1.9302 & 1.7074 & 2.00 & 28.12$^{+0.76}_{-0.74}$ &  3.310$^{+0.055}_{-0.055}$ & 27.07$\pm$0.15 & 3.216$\pm$0.013 \\
9970396  & 0.0086 $\pm$ 0.0004 & 0.2642 $\pm$ 0.0008 & 5.03$^{+0.23}_{-0.23}$ & 5.11  & 2.0245 & 1.8763 & 2.00 & 62.49$^{+1.12}_{-1.12}$ &  6.260$^{+0.077}_{-0.075}$ & 63.70$\pm$0.16 & 6.320$\pm$0.010 \\
10001167 & 0.0092 $\pm$ 0.0002 & 0.2655  &19.81  & 19.99  & > 4    & 2.4384 & 2.00 & ---   &   ---  & 19.90$\pm$0.09 & 2.762$\pm$0.012      \\
{ 10001167T} & 0.0050 $\pm$ 0.0006 & 0.2570 $\pm$ 0.0004 &10.08$^{+0.44}_{-0.44}$ & 10.20  & 1.9057    & 1.8568 & 2.00 & 18.95$^{+0.28}_{-0.27}$ &  2.754$^{+0.030}_{-0.030}$  & 19.90$\pm$0.09 & 2.762$\pm$0.012      \\
        \hline
    \end{tabular}
    \label{tab:Yol12_t9}
\end{table*}

\begin{figure}
    \centering
\includegraphics[width=1.15\linewidth]{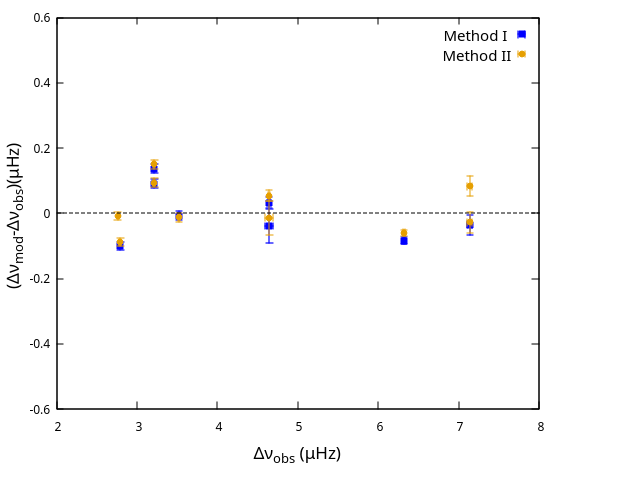}
\includegraphics[width=1.15\linewidth]{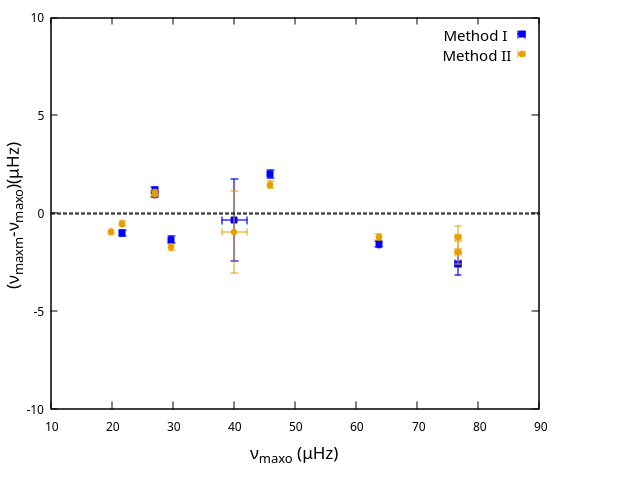}
    \caption{
    (a) Difference between the modelled and observed $\Delta\nu$ values for the SLO red giants.  
    (b) Difference between the modelled and observed $\nu_{\rm max}$ values. Squares and circles represent Methods I and II, respectively. Except for the Method I solution of KIC 8430105, all systems show agreement within approximately $2~\mu{\rm Hz}$.
    }
    \label{fig:Dnunumax}
\end{figure}

For 7 out of the 8 systems analysed with Method II, the calibrated mixing-length parameter of the primary component is larger than that of the secondary component:
${\alpha_{\rm A} > \alpha_{\rm B}.}$

The results cluster around
${\alpha_{\rm A}\approx 2,
\quad
\alpha_{\rm A}\approx \alpha_{\rm B}+0.18,}$
suggesting a possible dependence of $\alpha$ on evolutionary state. { Such a trend is less evident in Method I because of the larger scatter in the derived chemical compositions.}

The main source of uncertainty in RC models is the unknown amount of mass loss experienced before the helium-burning phase \citep{2025MNRAS.544..181O}. In the present work, we adopt the simplifying assumption of constant mass evolution. More realistic modelling of RC stars should also account for possible mass accretion by the secondary component. { These issues are beyond the scope of the present paper and will be investigated separately.}

The asteroseismic properties predicted by the models are compared with the observations in Fig. \ref{fig:Dnunumax}. Method I provides acceptable solutions for seven systems, whereas Method II yields solutions for all eight systems. The a panel shows that the differences between modelled and observed $\Delta\nu$ values are generally smaller than $0.15~\mu{\rm Hz}$. The b panel shows that the differences in $\nu_{\rm max}$ are smaller than $2~\mu{\rm Hz}$ for all systems except KIC 8430105 in Method I. Overall, both methods provide very similar asteroseismic results, with Method II producing slightly more consistent solutions.

{ Although the absolute differences between the observed and modelled $\Delta\nu$ and $\nu_{\rm max}$ values are small, in some cases they exceed the formal observational uncertainties. Several factors may contribute to this. First, different observational studies may report $\Delta\nu$ values that differ by more than the quoted uncertainties. Second, uncertainties in the dynamical masses and radii propagate into the model-derived $\Delta\nu$ and $\nu_{\rm max}$ values. Third, the determination of a mean $\Delta\nu$ depends on the adopted frequency range. This is particularly important for evolved RG stars, where only a limited number of detectable radial modes are available in both the observations and the models, introducing an additional source of uncertainty.}

In Table \ref{tab:Yol12_t9}, the fitting-formula ages $t_{\rm f}$ computed from equation (\ref{eq:A2}) are also listed. The agreement between $t_{\rm f}$ and the ages derived from Method II is generally excellent, supporting the internal consistency of the modelling approach.

The uncertainties listed in Table \ref{tab:Yol12_t9} are estimated using several complementary methods. The uncertainties in $Z_0$ are derived from the observational [Fe/H] uncertainties through
$Z = 10^{\rm [Fe/H]} Z_{\sun}.$

The uncertainties in $Y_0$ are estimated by propagating the corresponding uncertainty in $Z_0$ through the effective $c_{YZ}$ coefficient. The mean $c_{YZ}$ values are approximately 2.7 for Method I and 2.0 for Method II.

The uncertainties in $\Delta\nu$ and $\nu_{\rm max}$ are estimated using Monte Carlo simulations based on the observational uncertainties in $M$, $R$, and $T_{\rm eff}$. For each system, 1000 synthetic realizations are generated assuming Gaussian distributions for the observational parameters. The standard deviations of the resulting $\Delta\nu$ and $\nu_{\rm max}$ distributions are adopted as the model uncertainties. We also applied an MCMC approach, which yielded uncertainties approximately 30 per cent smaller than those from the Monte Carlo simulations.

The age uncertainties are estimated consistently using equation (\ref{eq:A3}) for each Monte Carlo realization. The resulting distributions provide the median age and the corresponding $1\sigma$ confidence intervals. The mean relative uncertainty in age is approximately 11 per cent. For comparison, the analytical age--mass relation
$t \propto M^{-3.2}$
yields a typical uncertainty of about 13 per cent, slightly larger than the uncertainty derived from the calibrated fitting relation.

\subsection{Chemical evolution}

The asteroseismic and non-asteroseismic model results obtained from the best-fitting solutions of the EBs using Methods I and II are listed in Table \ref{tab:Yol12_t9}. 
{ The derived initial helium and metallicity values provide additional constraints on the chemical evolution of the systems.}

\begin{figure}
    \centering
\includegraphics[width=1.15\linewidth]{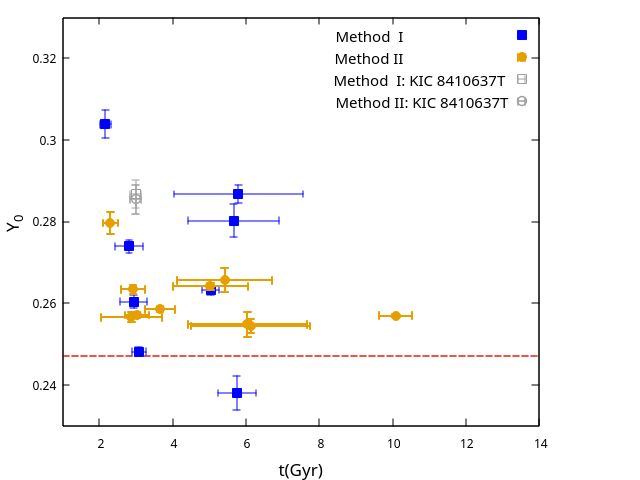}    
\includegraphics[width=1.15\linewidth]{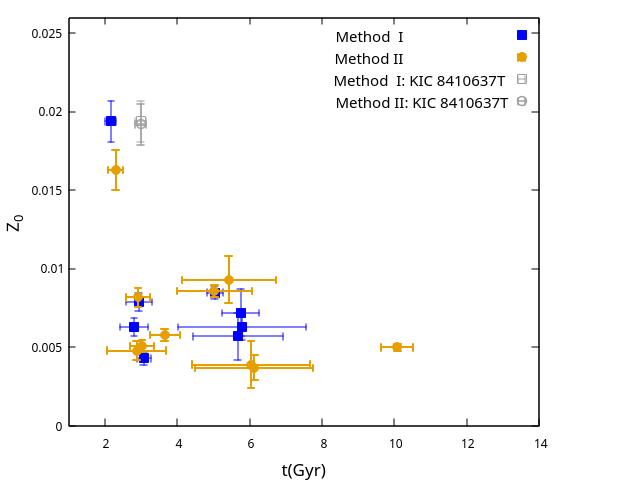}
    \caption{
    (a) The values of $Y_0$ obtained from Method I (filled squares) and Method II (filled circles) are plotted against $t$. The primordial helium abundance, $Y_{\rm p}=0.2471$ (Planck Collaboration et al. 2020), is shown by the dotted line. (b) $Z_0$ is plotted against $t$. The empty square and circle show the alternative solution for KIC 8410637.
    }
    \label{fig:ZtYt}
\end{figure}

In the upper and lower panels of Fig. \ref{fig:ZtYt}, the $Y_0$ and $Z_{0}$ values obtained from Methods I and II are plotted as functions of age, respectively. 
For almost all compatible solutions, the derived $Y_0$ values are greater than the primordial helium abundance $Y_{\rm p}$. 
Among the eight systems, only KIC 7037405 yields a sub-primordial helium abundance in Method I ($Y_{0}=0.2381 < Y_{\rm p}$). 
When the observed metallicity reported by \cite{2018MNRAS.476.3729B} is adopted for this system, the models predict luminosities slightly higher than the observed values. 
To reproduce the observed position in the HRD, the helium abundance must therefore be reduced, leading to a formally sub-primordial $Y_0$ value. 
This result most likely reflects limitations in the adopted input physics or uncertainties in the observational parameters rather than a physically meaningful helium abundance. 
In contrast, Method II yields more plausible parameters for this system, namely $Z_0 = 0.0093$ and $Y_0 = 0.2657$.

Among the Method I solutions, the lowest metallicity is obtained for KIC 8430105 ($Z_0 = 0.0043$), whereas Method II gives the lowest value for KIC 7377422 ($Z_0 = 0.0037$). 
The overall spread in both $Y_0$ and $Z_0$ increases towards younger systems, { indicating a larger diversity in the chemical properties of relatively young binaries.}

The grey open circle and open square in Fig. \ref{fig:ZtYt} correspond to the alternative solution for KIC 8410637. 
The values of $Y_0$, $Z_0$, and age derived from the two methods are mutually consistent for this solution. 
Overall, the distributions shown in Fig. \ref{fig:ZtYt} illustrate the temporal behaviour of helium and metal enrichment among the analysed EBs.

\subsection{Asteroseismic analysis}

We calculate the adiabatic oscillation frequencies of the interior models that satisfy the non-asteroseismic observational constraints using the {\small ADIPLS} package integrated into {\small MESA}. 
The corresponding $\Delta\nu_{\rm mod}$ values are computed directly from the model frequencies. 
We also calculate $\nu_{\rm max}$ from the model values of $\Gamma_{1}$, $\mu$, $g$, and $T_{\rm eff}$, while $\Delta\nu_{\rm sca}$ is derived from the mean density scaling relation. 
In general, the modelled and observed asteroseismic parameters are in good agreement (see Fig. \ref{fig:Dnunumax}).

\subsubsection{Role of the reference stars in the scaling relations}
\label{sec:461}
\begin{figure}
    \centering
\includegraphics[width=1.2\linewidth]{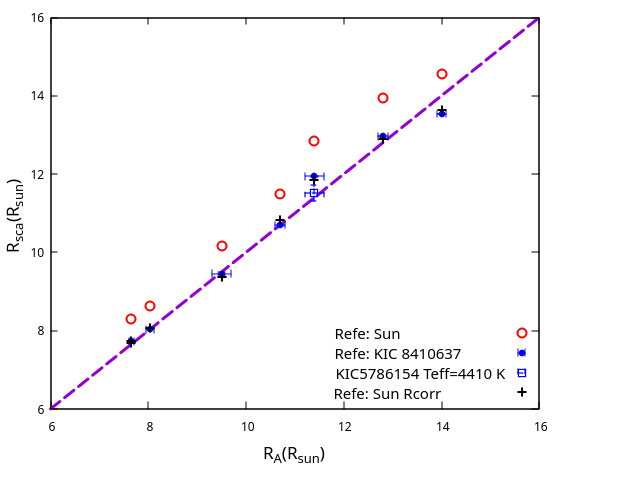}
\includegraphics[width=1.2\linewidth]{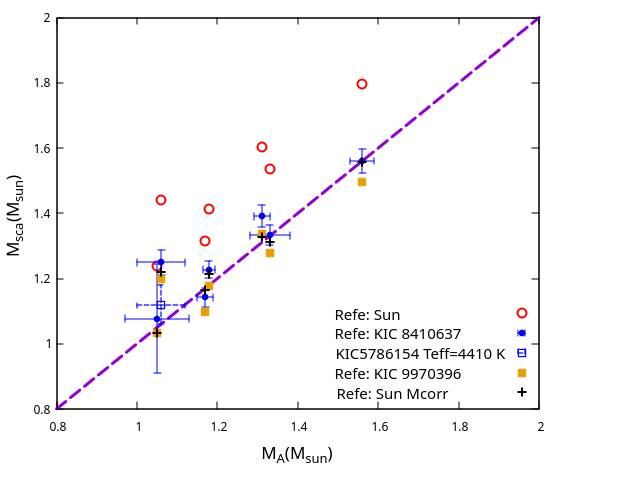}
    \caption{
   (a) $R_{\rm sca}$ computed using KIC 8410637 A as the reference star in the asteroseismic scaling relations plotted against the observed radii of the primary components (filled circles). 
    For comparison, the values obtained using the Sun as the reference star are also shown (open circles). 
    The squares correspond to the results obtained using KIC 9970396 A as the reference star. 
    { The plus symbols represent the corrected radii ($R'_{\rm sca}$) obtained from the non-standard scaling relations using the solar reference together with the correction factors $f_{\nu_{\rm max}}$ and $f_{\Delta \nu}$.
    (b) Comparison between the scaling-relation masses ($M_{\rm sca}$) and the observed dynamical masses. 
    The plus symbols represent the corrected masses ($M'_{\rm sca}$) obtained from the non-standard scaling relations using the solar reference together with the correction factors $f_{\nu_{\rm max}}$ and $f_{\Delta \nu}$.}
    }
    \label{fig:RscaMsca}
\end{figure}

There are significant differences between the dynamical masses and radii of these EBs and the values obtained from the classical asteroseismic scaling relations (equation \ref{eq:clasicsca}). 
These discrepancies indicate that corrections to the scaling relations are required. 
Many studies have investigated this problem \citep{2017ApJ...844..102H,2022ApJ...927..167L,2023MNRAS.518.5552Y}. 

In this study, instead of adopting the Sun as the reference star, we use one of the binaries with good agreement between models and observations, namely KIC 8410637 A, as a reference star for the scaling relations. 
In the upper panel of Fig. \ref{fig:RscaMsca}, the scaling-relation radii obtained using KIC 8410637 A are compared with the observed radii of the primary stars. 
For comparison, the radii obtained using the Sun as the reference star are also shown. 
The agreement between $R_{\rm sca}$ and the observed radii is substantially improved when KIC 8410637 A is used instead of the Sun. 

{ We also computed the non-standard scaling-relation radii ($R'_{\rm sca}$; equation \ref{eq:non-standardRsca}) using the Sun as the reference star together with the correction factors $f_{\nu_{\rm max}}$ from \cite{2023MNRAS.518.5552Y} and $f_{\Delta \nu}$ from \cite{Sharma}. These corrected radii, shown with plus symbols in Fig. \ref{fig:RscaMsca}(a), are also in very good agreement with the observed radii and with the results obtained using KIC 8410637 A as the reference star.}

We also compute the scaling-relation radii using KIC 9970396 A as the reference star and obtain very similar results.

The lower panel of Fig. \ref{fig:RscaMsca} compares the scaling-relation masses with the dynamical masses. 
The classical scaling relations based on the Sun produce systematically larger discrepancies, whereas the use of KIC 8410637 A and KIC 9970396 A as reference stars significantly improves the agreement. 
{ We also computed the non-standard scaling-relation masses ($M'_{\rm sca}$; equation \ref{eq:non-standardMsca}) using the solar reference together with the correction factors $f_{\nu_{\rm max}}$ from \cite{2023MNRAS.518.5552Y} and $f_{\Delta \nu}$ from \cite{Sharma}. The corrected masses, shown with plus symbols in Fig. \ref{fig:RscaMsca}(b), are in good agreement with the dynamical masses and are comparable to the results obtained using KIC 8410637 A as the reference star.} 
Among these, KIC 8410637 A provides the closest consistency between $M_{\rm sca}$ and the dynamical masses.

The comparison between dynamical and asteroseismic masses and radii reveals the largest discrepancy for KIC 5786154. 
This difference may partly originate from uncertainties in $T_{\rm eff}$. 
\cite{2016ApJ...832..121G} reported the highest effective temperature for this system ($T_{\rm eff}=4747$ K). 
We therefore explored the temperature required for consistency between the asteroseismic and dynamical radii. 
The points with error bars in Fig. \ref{fig:RscaMsca} correspond to the result obtained for $T_{\rm eff}=4410$ K.

\subsubsection{Comparison of the model and observed individual frequencies of Tek Ayak (KIC 8410637)}

Among the systems analysed in this study, individual oscillation frequencies are available for two stars from the {\it Kepler} light curves. 
One of these stars is Tek Ayak (KIC 8410637), which is also adopted as the reference star in the modified scaling relations discussed above. 

Interior models of Tek Ayak are constructed with the {\small MESA} astero module in order to compare the observed and modelled oscillation frequencies. 
The astero module uses two separate inlists. 
The stellar evolution calculations are controlled through {\tt inlist\_example\_astero}, while the observational constraints are specified in {\tt inlist\_astero\_search\_controls}. 
These constraints include $T_{\rm eff}$, $R$, [M/H], $\Delta\nu$, $\nu_{\rm max}$, and the individual oscillation frequencies. 
The oscillation frequencies are computed using the {\small ADIPLS} package. 
Near-surface effects are corrected using the `combined' prescription of \cite{2014A&A...568A.123B}.

The surface correction adopted in the astero module is written as

\begin{equation}
\nu_{\rm diff}=
\frac{
a_{-1}(\nu/\nu_{\rm ac})^{-1}
+
a_{3}(\nu/\nu_{\rm ac})^{3}
}{I},
\end{equation}

where $\nu$ is the oscillation frequency, $\nu_{\rm ac}$ is the acoustic cut-off frequency, $I$ is the mode inertia, and $a_{-1}$ and $a_{3}$ are the correction coefficients. 
The acoustic cut-off frequency is computed from

\begin{equation}
\nu_{\rm ac}=
\nu_{\rm ac\sun}
\frac{g/g_{\sun}}
{\sqrt{T_{\rm eff}/T_{\rm eff\sun}}},
\end{equation}
where $\nu_{\rm ac\sun}$, $g_{\sun}$, and $T_{\rm eff\sun}$ are the solar reference values. {  $\nu_{\rm ac\sun}$ is taken as 5000 $\mu$Hz \citep{2014A&A...568A.123B}}. 

The fitted correction coefficients are 
$a_{-1}=-2.4374\times10^{-5}$ $\mu$Hz and 
$a_{3}=-2.303\times10^{-6}$ $\mu$Hz for Method I, and 
$a_{-1}=-2.626\times10^{-5}$ $\mu$Hz and 
$a_{3}=-5.085\times10^{-6}$ $\mu$Hz for Method II.

\begin{figure}
    \centering
\includegraphics[width=1.17\linewidth]{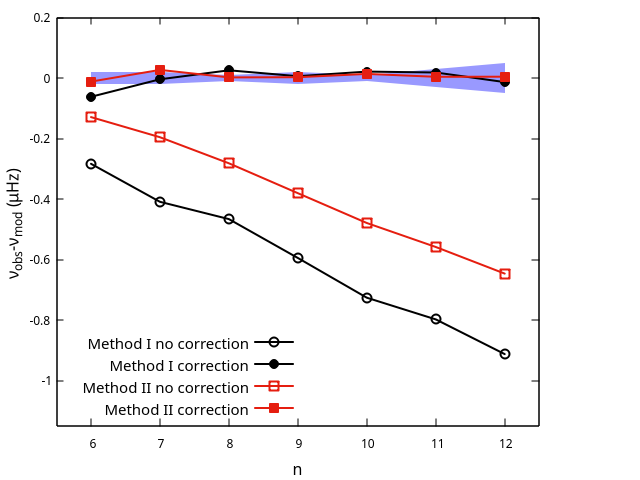}    
    \caption{
    Difference between observed and model frequencies, $\nu_{\rm obs}-\nu_{\rm mod}$, plotted against radial order $n$ for Tek Ayak (KIC 8410637). 
    The observed frequencies are from \protect\cite{2018MNRAS.478.4669T}. 
    Circles and squares represent the results obtained with Methods I and II, respectively. 
    Open symbols show the uncorrected frequencies, while filled symbols show the frequencies after applying the surface correction. 
    The shaded region indicates the observational uncertainty range for the radial ($l=0$) modes.
    }
    \label{fig:correction_nu}
\end{figure}

Fig. \ref{fig:correction_nu} illustrates the effect of the surface correction on the $l=0$ frequencies. 
The correction becomes larger at higher frequencies. 
For Method I, the correction ranges from 0.28 to 0.91 $\mu$Hz, whereas for Method II it ranges from 0.13 to 0.65 $\mu$Hz. 
After the correction is applied, both methods reproduce the observed frequencies very well. 
In Method I, however, the frequencies corresponding to $n=6$ and 8 remain slightly outside the observational error range.

The interior models used in the astero calculations are based on the parameters listed in Table \ref{tab:Yol12_t9}. 
The resulting adiabatic oscillation frequencies are listed in Table \ref{tab:astero_freq}. 

For Method I, the derived global parameters are:
$\Delta\nu = 4.5735$ $\mu$Hz,
$\nu_{\rm max} = 46.5033$ $\mu$Hz,
$T_{\rm eff}=4800.8$ K,
$R=10.68~R_{\sun}$,
and $t=2.167$ Gyr.

For Method II, the corresponding values are:
$\Delta\nu = 4.5790$ $\mu$Hz,
$\nu_{\rm max} = 46.2160$ $\mu$Hz,
$T_{\rm eff}=4799.5$ K,
$R=10.71~R_{\sun}$,
and $t=2.303$ Gyr.

The modelled $\Delta\nu$ values are derived from the $\Delta\nu-\nu$ relation using the $l=0$ frequencies. 
Applying the same procedure to the observed frequencies yields 
$\Delta\nu_{\rm obs}=4.5817\pm0.0577$ $\mu$Hz. 
The model values are therefore in very good agreement with the observations. 
The observed $\nu_{\rm max}$ value is reproduced slightly better by Method II.

\begin{table}
    \centering
    \caption{
    Model and observed oscillation frequencies of Tek Ayak (KIC 8410637). 
    The columns list the mode degree $l$, the model frequencies obtained from Methods I and II, and the observed frequencies. 
    The final rows give the corresponding $\nu_{\rm max}$ and $\Delta\nu$ values. 
    Observational frequencies are from \protect\cite{2018MNRAS.478.4669T}.
    }
    \begin{tabular}{ccccc}
        \hline
     & Method I & Method II &  \\
 $l$ & $\nu$ ($\mu$Hz) & $\nu$ ($\mu$Hz) & $\nu_{\rm obs}$ ($\mu$Hz) \\
        \hline
0  & 32.85 & 32.80 & 32.79$\pm$0.02 \\
0  & 37.18 & 37.15 & 37.18$\pm$0.02 \\
0  & 41.59 & 41.62 & 41.62$\pm$0.01 \\
0  & 46.27 & 46.28 & 46.28$\pm$0.02 \\
0  & 50.83 & 50.84 & 50.85$\pm$0.01 \\
0  & 55.52 & 55.54 & 55.54$\pm$0.03 \\
0  & 60.29 & 60.28 & 60.28$\pm$0.05 \\
1  & 35.10 & 35.45 & 35.16$\pm$0.04 \\
1  & 39.45 & 39.37 & 39.44$\pm$0.02 \\
1  & 44.07 & 44.05 & 44.06$\pm$0.02 \\
1  & 49.00 & 48.97 & 48.70$\pm$0.02 \\
1  & 53.78 & 52.97 & 53.32$\pm$0.02 \\
1  & 58.43 & 57.92 & 58.04$\pm$0.04 \\
1  & 63.98 & 62.91 & 62.95$\pm$0.02 \\
2  & 31.14 & 32.16 & 32.13$\pm$0.05 \\
2  & 36.54 & 36.50 & 36.56$\pm$0.03 \\
2  & 41.02 & 41.06 & 41.02$\pm$0.01 \\
2  & 45.63 & 45.58 & 45.69$\pm$0.01 \\
2  & 50.24 & 50.50 & 50.31$\pm$0.02 \\
2  & 54.99 & 55.10 & 55.06$\pm$0.04 \\
2  & 60.46 & 60.01 & 59.74$\pm$0.05 \\
3  & 37.76 & 37.71 & 37.70$\pm$0.09 \\
3  & 47.37 & 47.02 & 47.19$\pm$0.01 \\
3  & 51.02 & 51.79 & 51.71$\pm$0.06 \\
\hline
$\nu_{\rm max}$ ($\mu$Hz) & 46.50 & 46.22 & 46.00$\pm$0.19 \\
$\Delta\nu$ ($\mu$Hz) & 4.574 & 4.579 & 4.641$\pm$0.017 \\
$\Delta\nu$ ($\mu$Hz) & --- & --- & 4.5817$\pm$0.0577 \\
\hline
\end{tabular}
\label{tab:astero_freq}
\end{table}

The differences between the observed and model frequencies are smaller than 1 per cent for both methods. 
For Method I, the mean fractional differences for the modes with $l=0$, 1, 2, and 3 are 0.0001, 0.0030, 0.0011, and 0.0039, respectively. 
For Method II, the corresponding values are $-0.0001$, 0.0004, 0.0010, and $-0.0008$. 

Although both methods reproduce the individual frequencies well, the detailed pattern in the $\Delta\nu-\nu$ relation differs slightly. 
The two models are similarly successful for the $l=0$ frequencies, but Method II reproduces the observed pattern more accurately overall.

The second system for which individual frequencies are available is KIC 9540226 \citep{2018MNRAS.478.4669T}. 
Although the agreement is not as good as that obtained for Tek Ayak, interior models were also constructed for this system with the astero module. 
For Method I, the mean fractional differences between model and observed frequencies for $l=0$, 1, and 2 are 0.0003, $-0.0008$, and 0.0037, respectively. 
For Method II, the corresponding values are 0.0001, $-0.0011$, and 0.0024. 

The $\Delta\nu$ values derived from the $l=0$ frequencies are 3.165 and 3.169 $\mu$Hz for Methods I and II, respectively, while the observational value obtained from the $\Delta\nu-\nu$ diagram is 3.194 $\mu$Hz. 
The agreement between the models and observations is therefore weaker than in the case of Tek Ayak. 
The corresponding $\nu_{\rm max}$ values are 28.09 and 27.60 $\mu$Hz for Methods I and II, respectively.

\section{Conclusions}
\label{sec:sec5}

{  In this study, we analysed 11 double-lined EBs containing solar-like oscillators using two { complementary modelling approaches based on detailed stellar evolution calculations performed with {\small MESA}  
\citep{2013A&A...556A.138F,2016ApJ...832..121G,2016ApJ...818..108R,2018MNRAS.476.3729B,2022A&A...668A..82B,2018MNRAS.478.4669T,2019MNRAS.484..451H,2021A&A...648A.113B,2022MNRAS.517.4187T,2025A&A...699A.152T}.
The models are constructed individually for the component stars by using the observed masses and trial chemical compositions, and the ages are determined from the coevality condition of the binary components. The main results of this study can be summarized as follows.}

\begin{enumerate}

\item 
{ Method I determines the age and initial helium abundance of the systems by adopting the observed metallicity and varying $Y_0$ until the luminosity calibration of the two components yields the same age. }
For the systems with acceptable solutions, Method I gives ages in the range 2.17$-$5.79 Gyr and initial helium abundances in the range $Y_0=0.2381$-$0.3039$. 
{ The age difference between the component models varies almost linearly with $Y_0$, which allows a straightforward determination of the coeval solution.}

\item
{ Method II determines the age and initial metallicity simultaneously by assuming a chemical enrichment relation between $Y_0$ and $Z_0$, adopting $Y_0=Y_{\rm p}+2Z_0$. 
In this case, the coevality condition is satisfied by varying $Z_0$. 
Method II yields ages between 2.30 and 10.08 Gyr and a narrower range of helium abundances, $Y_0=0.2545$-$0.2797$. 
All solutions obtained with Method II are physically consistent with $Y_0 > Y_{\rm p}$.}

\item
The comparison between the two methods indicates that Method II generally provides more stable and physically plausible solutions, especially for systems with uncertain spectroscopic metallicities. 
The results also show that the inferred ages and chemical compositions can depend significantly on the adopted observational data set.

\item
{ The evolutionary tracks constructed using the observed masses reproduce the observed luminosities of most systems successfully. 
However, agreement in the HRD alone is not sufficient for a physically acceptable solution. 
The simultaneous requirement of coevality provides a much stronger constraint on the stellar models.}

\item
{ The modelled asteroseismic parameters are in good agreement with the observations. 
For almost all systems, the differences between model and observed $\Delta\nu$ values are smaller than 0.15 $\mu$Hz, while the differences in $\nu_{\rm max}$ are generally below 2 $\mu$Hz. 
The ages obtained from the detailed interior models are also in excellent agreement with the fitting formula derived from the model grids.}

\item
The comparison between dynamical and asteroseismic masses and radii demonstrates that the classical scaling relations can be improved significantly by using evolved solar-like oscillators such as KIC 8410637 and KIC 9970396 as reference stars instead of the Sun. { The agreement between $M_{\rm sca}$, $R_{\rm sca}$ and the dynamical values becomes substantially better when these evolved stars are adopted as reference stars. We also show that the non-standard scaling relations, combined with the correction factors $f_{\nu_{\rm max}}$ and $f_{\Delta \nu}$ while retaining the Sun as the reference star, provide similarly improved agreement with the dynamical masses and radii.

\item
For Tek Ayak (KIC 8410637), detailed frequency modelling performed with the {\small MESA} astero module yields oscillation frequencies that agree very well with the observed frequencies from {\it Kepler}. 
After surface corrections are applied, the differences between model and observed frequencies remain below approximately 1 per cent. 
Both Methods I and II reproduce the observed oscillation spectrum successfully, although Method II provides slightly better agreement in the $\Delta\nu-\nu$ pattern.}

\item
The inferred age--metallicity and age--helium relations are consistent with Galactic chemical evolution. 
Older systems tend to have lower metallicities and lower helium abundances, while younger systems show larger spreads in both parameters. 
The results obtained with Method II are broadly consistent with the chemical enrichment relation $Y_0 \approx Y_{\rm p}+2Z_0$.

\end{enumerate}

{ Overall, this study demonstrates that eclipsing binaries containing solar-like oscillators provide exceptionally strong constraints on stellar interior models because masses, radii, luminosities, effective temperatures, and asteroseismic properties can be analysed simultaneously. 
The combination of binary constraints and asteroseismology enables detailed tests of stellar evolution, chemical enrichment, and asteroseismic scaling relations. 
Future studies based on improved spectroscopic abundances and more precise oscillation frequencies will further strengthen these constraints and help refine the physics of stellar structure and evolution.}

\section*{Acknowledgements}
This study was supported by Scientific and Technological Research Council of Turkey (T\"UBITAK) under the Grant Number 123F019. We are grateful to Ege University Planning and Monitoring Coordination of Organizational Development and Directorate of Library and Documentation for their support in the editing and proofreading service of this study.

\section*{Data Availability}

 The data underlying this article will be shared on reasonable request to the corresponding author.




\begin{thebibliography}{99}
\bibitem[\protect\citeauthoryear{Abdurro'uf et al.}{2022}]{2022ApJS..259...35A} Abdurro'uf, Accetta K., Aerts C., Silva Aguirre V., Ahumada R., Ajgaonkar N., Filiz Ak N., et al., 2022, ApJS, 259, 35. doi:10.3847/1538-4365/ac4414
\bibitem[\protect\citeauthoryear{Asplund et al.}{2009}]{Asplund2009} Asplund M., Grevesse N., Sauval A.~J., Scott P., 2009, ARA\&A, 47, 481. doi:10.1146/annurev.astro.46.060407.145222
\bibitem[\protect\citeauthoryear{Ball \& Gizon}{2014}]{2014A&A...568A.123B} {Ball W.~H., Gizon L., 2014, A\&A, 568, A123. doi:10.1051/0004-6361/201424325}
\bibitem[\protect\citeauthoryear{Beck et al.}{2022}]{2022A&A...667A..31B} Beck P.~G., Mathur S., Hambleton K., Garc{\'\i}a R.~A., Steinwender L., Eisner N.~L., do Nascimento J.-D., et al., 2022, A\&A, 667, A31. doi:10.1051/0004-6361/202143005
\bibitem[\protect\citeauthoryear{Beck et al.}{2024}]{2024A&A...682A...7B} Beck P.~G., Grossmann D.~H., Steinwender L., Schimak L.~S., Muntean N., Vrard M., Patton R.~A., et al., 2024, A\&A, 682, A7. doi:10.1051/0004-6361/202346810
\bibitem[\protect\citeauthoryear{Benbakoura et al.}{2021}]{2021A&A...648A.113B} Benbakoura M., Gaulme P., McKeever J., Sekaran S., Beck P.~G., Spada F., Jackiewicz J., et al., 2021, A\&A, 648, A113. doi:10.1051/0004-6361/202037783
\bibitem[\protect\citeauthoryear{B{\"{o}}hm-Vitense et al.}{1958}]{Bohm1958}
B{\"{o}}hm-Vitense E., 1958, Zeitschrift f{\"{u}}r Astrophysik, 46, 108
\bibitem[\protect\citeauthoryear{Borucki et al.}{2010}]{Borucki2010} Borucki W. J., Koch D., Basri G., Batalha, N., Broen, T., et al. 2010, Science, 327, 977
\bibitem[\protect\citeauthoryear{Brogaard et al.}{2016}]{2016AN....337..793B} Brogaard K., Jessen-Hansen J., Handberg R., Arentoft T., Frandsen S., Grundahl F., Bruntt H., et al., 2016, AN, 337, 793. doi:10.1002/asna.201612374
\bibitem[\protect\citeauthoryear{Brogaard et al.}{2018}]{2018MNRAS.476.3729B} Brogaard K., Hansen C.~J., Miglio A., Slumstrup D., Frandsen S., Jessen-Hansen J., Lund M.~N., et al., 2018, MNRAS, 476, 3729. doi:10.1093/mnras/sty268
\bibitem[\protect\citeauthoryear{Brogaard et al.}{2022}]{2022A&A...668A..82B} Brogaard K., Arentoft T., Slumstrup D., Grundahl F., Lund M.~N., Arndt L., Grund S., et al., 2022, A\&A, 668, A82. doi:10.1051/0004-6361/202244345
\bibitem[\protect\citeauthoryear{Brown et al.}{1991}]{Brown1991} Brown, T.M., Gilliland R.L., Noyes, R.W., Ramsey, L.W., 1991, ApJ, 368, 599
\bibitem[\protect\citeauthoryear{Brown et al.}{2011}]{2011AJ....142..112B} Brown T.~M., Latham D.~W., Everett M.~E., Esquerdo G.~A., 2011, AJ, 142, 112. doi:10.1088/0004-6256/142/4/112
\bibitem[\protect\citeauthoryear{Casagrande et al.}{2007}]{2007MNRAS.382.1516C} Casagrande L., Flynn C., Portinari L., Girardi L., Jimenez R., 2007, MNRAS, 382, 1516. doi:10.1111/j.1365-2966.2007.12512.x
{{\bibitem[\protect\citeauthoryear{Castelli \& Kurucz}{2003}]{2003IAUS..210P.A20C} Castelli F., Kurucz R.~L., 2003, IAUS, 210, A20. doi:10.48550/arXiv.astro-ph/0405087}}
\bibitem[\protect\citeauthoryear{Choi et al.}{2016}]{2016ApJ...823..102C} Choi J., Dotter A., Conroy C., Cantiello M., Paxton B., Johnson B.~D., 2016, ApJ, 823, 102. doi:10.3847/0004-637X/823/2/102
\bibitem[\protect\citeauthoryear{Christensen-Dalsgaard}{1993}]{1993ASPC...42..347C} Christensen-Dalsgaard J., 1993, ASPC, 42, 347
\bibitem[\protect\citeauthoryear{Dotter}{2016}]{2016ApJS..222....8D} Dotter A., 2016, ApJS, 222, 8. doi:10.3847/0067-0049/222/1/8
\bibitem[\protect\citeauthoryear{Ferguson et al.}{2005}]{Ferguson2005} Ferguson, J.W., Alexander, D.R., Allard, F., Barmanu, T., et al. 2005, ApJ, 623, 585
\bibitem[\protect\citeauthoryear{Frandsen et al.}{2013}]{2013A&A...556A.138F} Frandsen S., Lehmann H., Hekker S., Southworth J., Debosscher J., Beck P., Hartmann M., et al., 2013, A\&A, 556, A138. doi:10.1051/0004-6361/201321817
\bibitem[\protect\citeauthoryear{Frasca et al.}{2016}]{2016AA...594A..39F} Frasca A., Molenda-{\.Z}akowicz J., De Cat P., Catanzaro G., Fu J.~N., Ren A.~B., Luo A.~L., et al., 2016, A\&A, 594, A39. doi:10.1051/0004-6361/201628337
\bibitem[\protect\citeauthoryear{Gaia Collaboration et al.}{2018}]{Gaia2018} Gaia Collaboration, Brown A. G. A., Vallenari, A., Prusti, T., de Bruijne, J.H.J., et al., 2018, A\&A, 616, A1
\bibitem[\protect\citeauthoryear{Gaia Collaboration et al.}{2021}]{Gaia2021} Gaia Collaboration, Brown A. G. A., Vallenari, A., Prusti, T., de Bruijne, J.H.J., et al., 2021, A\&A, 649, A1
\bibitem[\protect\citeauthoryear{Gaulme et al.}{2016}]{2016ApJ...832..121G} Gaulme P., McKeever J., Jackiewicz J., Rawls M.~L., Corsaro E., Mosser B., Southworth J., et al., 2016, ApJ, 832, 121. doi:10.3847/0004-637X/832/2/121
\bibitem[\protect\citeauthoryear{Gaulme et al.}{2020}]{2020A&A...639A..63G} Gaulme P., Jackiewicz J., Spada F., Chojnowski D., Mosser B., McKeever J., Hedlund A., et al., 2020, A\&A, 639, A63. doi:10.1051/0004-6361/202037781
\bibitem[\protect\citeauthoryear{Grossmann et al.}{2025}]{2025arXiv250109018G} Grossmann D.~H., Beck P.~G., Mathur S., Johnston C., Godoy-Rivera D., Zinn J.~C., Cassisi S., et al., 2025, arXiv, arXiv:2501.09018. doi:10.48550/arXiv.2501.09018
{{\bibitem[\protect\citeauthoryear{Hauschildt, Allard, \& Baron}{1999}]{1999ApJ...512..377H} Hauschildt P.~H., Allard F., Baron E., 1999, ApJ, 512, 377. doi:10.1086/306745
\bibitem[\protect\citeauthoryear{Hauschildt et al.}{1999}]{1999ApJ...525..871H} Hauschildt P.~H., Allard F., Ferguson J., Baron E., Alexander D.~R., 1999, ApJ, 525, 871. doi:10.1086/307954}}
\bibitem[\protect\citeauthoryear{Hekker et al.}{2010}]{2010ApJ...713L.187H} Hekker S., Debosscher J., Huber D., Hidas M.~G., De Ridder J., Aerts C., Stello D., et al., 2010, ApJL, 713, L187. doi:10.1088/2041-8205/713/2/L187
{\bibitem[\protect\citeauthoryear{Helminiak et al.}{2019}]{2019MNRAS.484..451H} Helminiak K.~G., Konacki M., Maehara H., Kambe E., Ukita N., Ratajczak M., Pigulski A., et al., 2019, MNRAS, 484, 451. doi:10.1093/mnras/sty3528}
\bibitem[\protect\citeauthoryear{Huber et al.}{2017}]{2017ApJ...844..102H} Huber D., Zinn J., Bojsen-Hansen M., Pinsonneault M., Sahlholdt C., Serenelli A., Silva Aguirre V., et al., 2017, ApJ, 844, 102. doi:10.3847/1538-4357/aa75ca
\bibitem[\protect\citeauthoryear{Iglesias \& Rogers}{1993}]{Iglesias1993} Iglesias C.~A., Rogers F.~J., 1993, ApJ, 412, 752. doi:10.1086/172958
\bibitem[\protect\citeauthoryear{Iglesias \& Rogers}{1996}]{Iglesias1996} Iglesias C.A., and Rogers F.J. 1996, ApJ, 464, 943
\bibitem[\protect\citeauthoryear{Jermyn et al.}{2023}]{Jermyn2023} Jermyn A.~S., Bauer E.~B., Schwab J., Farmer R., Ball W.~H., Bellinger E.~P., Dotter A., et al., 2023, ApJS, 265, 15. doi:10.3847/1538-4365/acae8d
\bibitem[\protect\citeauthoryear{Kjeldsen \& Bedding }{1995}]{Kjebed1995} Kjeldsen H., and Bedding T. R. 1995, A\&A, 293, 87
\bibitem[\protect\citeauthoryear{Li et al.}{2022}]{2022ApJ...927..167L} Li T., Li Y., Bi S., Bedding T.~R., Davies G., Du M., 2022, ApJ, 927, 167. doi:10.3847/1538-4357/ac4fbf
\bibitem[\protect\citeauthoryear{Miglio et al.}{2021}]{2021A&A...645A..85M} Miglio A., Chiappini C., Mackereth J.~T., Davies G.~R., Brogaard K., Casagrande L., Chaplin W.~J., et al., 2021, A\&A, 645, A85. doi:10.1051/0004-6361/202038307
\bibitem[\protect\citeauthoryear{{\"O}rtel \& Y{\i}ld{\i}z}{2025}]{2025MNRAS.544..181O} {\"O}rtel S., Y{\i}ld{\i}z M., 2025, MNRAS, 544, 181. doi:10.1093/mnras/staf1695
\bibitem[\protect\citeauthoryear{Paquette et al.}{1986}]{1986ApJS...61..177P} Paquette C., Pelletier C., Fontaine G., Michaud G., 1986, ApJS, 61, 177. doi:10.1086/191111
\bibitem[\protect\citeauthoryear{Paxton et al.}{2011}]{Paxton2011} 
Paxton B., Bildsten L., Dotter A., Herwig F., Lesaffre P. and Timmes F., 2011, ApJS, 192, 35
\bibitem[\protect\citeauthoryear{Paxton et al.}{2013}]{Paxton2013} 
Paxton B., Cantiello M., Arras P., Bildsten L., Brown, E.F, et al., 2013, ApJS, 208, 42
\bibitem[\protect\citeauthoryear{Paxton et al.}{2015}]{Paxton2015} Paxton B., Marchant P., Schwab J., Bauer E.~B., Bildsten L., Cantiello M., Dessart L., et al., 2015, ApJS, 220, 15. doi:10.1088/0067-0049/220/1/15
\bibitem[\protect\citeauthoryear{Paxton et al.}{2018}]{Paxton2018} Paxton B., Schwab J., Bauer E.~B., Bildsten L., Blinnikov S., Duffell P., Farmer R., et al., 2018, ApJS, 234, 34. doi:10.3847/1538-4365/aaa5a8
\bibitem[\protect\citeauthoryear{Paxton et al.}{2019}]{Paxton2019} Paxton B., Smolec R., Schwab J., Gautschy A., Bildsten L., Cantiello M., Dotter A., et al., 2019, ApJS, 243, 10. doi:10.3847/1538-4365/ab2241
\bibitem[\protect\citeauthoryear{Pinsonneault et al.}{2018}]{2018ApJS..239...32P} Pinsonneault M.~H., Elsworth Y.~P., Tayar J., Serenelli A., Stello D., Zinn J., Mathur S., et al., 2018, ApJS, 239, 32. doi:10.3847/1538-4365/aaebfd
\bibitem[\protect\citeauthoryear{Pinsonneault et al.}{2025}]{2025ApJS..276...69P} Pinsonneault M.~H., Zinn J.~C., Tayar J., Serenelli A., Garc{\'\i}a R.~A., Mathur S., Vrard M., et al., 2025, ApJS, 276, 69. doi:10.3847/1538-4365/ad9fef
\bibitem[\protect\citeauthoryear{Planck Collaboration et al.}{2020}]{2020A&A...641A...6P} Planck Collaboration, Aghanim N., Akrami Y., Ashdown M., Aumont J., Baccigalupi C., Ballardini M., et al., 2020, A\&A, 641, A6. doi:10.1051/0004-6361/201833910
\bibitem[\protect\citeauthoryear{Qian et al.}{2018}]{2018ApJS..235....5Q} Qian S.-B., Zhang J., He J.-J., Zhu L.-Y., Zhao E.-G., Shi X.-D., Zhou X., et al., 2018, ApJS, 235, 5. doi:10.3847/1538-4365/aaa601
\bibitem[\protect\citeauthoryear{Rawls et al.}{2016}]{2016ApJ...818..108R} Rawls M.~L., Gaulme P., McKeever J., Jackiewicz J., Orosz J.~A., Corsaro E., Beck P.~G., et al., 2016, ApJ, 818, 108. doi:10.3847/0004-637X/818/2/108
\bibitem[\protect\citeauthoryear{Rowan et al.}{2024}]{2024arXiv240902983R} Rowan D.~M., Stanek K.~Z., Kochanek C.~S., Thompson T.~A., Jayasinghe T., Blaum J., Fulton B.~J., et al., 2024, arXiv, arXiv:2409.02983. doi:10.48550/arXiv.2409.02983
\bibitem[\protect\citeauthoryear{Sharma et al.}{2016}]{Sharma}{ Sharma S., Stello D., Bland-Hawthorn J. et al. 2016, ApJ, 822, 15}
\bibitem[\protect\citeauthoryear{Skrutskie et al.}{2006}]{2006AJ....131.1163S} Skrutskie M.~F., Cutri R.~M., Stiening R., Weinberg M.~D., Schneider S., Carpenter J.~M., Beichman C., et al., 2006, AJ, 131, 1163. doi:10.1086/498708
\bibitem[\protect\citeauthoryear{Sullivan et al.}{2015}]{Sullivan2015} 
Sullivan, P. W., Winn, J.N., Berta-Thompson, Z.K., Charbonneau, D., Deming, D., et al., 2015, ApJ, 809, 77
\bibitem[\protect\citeauthoryear{Tassoul}{1980}]{1980ApJS...43..469T} Tassoul M., 1980, ApJS, 43, 469. doi:10.1086/190678
\bibitem[\protect\citeauthoryear{Theme{\ss}l et al.}{2018}]{2018MNRAS.478.4669T} Theme{\ss}l N., Hekker S., Southworth J., Beck P.~G., Pavlovski K., Tkachenko A., Angelou G.~C., et al., 2018, MNRAS, 478, 4669. doi:10.1093/mnras/sty1113
{\bibitem[\protect\citeauthoryear{Thomsen et al.}{2022}]{2022MNRAS.517.4187T} Thomsen J.~S., Brogaard K., Arentoft T., Slumstrup D., Lund M.~N., Grundahl F., Miglio A., et al., 2022, MNRAS, 517, 4187. doi:10.1093/mnras/stac2942
\bibitem[\protect\citeauthoryear{Thomsen et al.}{2025}]{2025A&A...699A.152T} Thomsen J.~S., Miglio A., Brogaard K., Montalb{\'a}n J., Tailo M., van Rossem W.~E., Casali G., et al., 2025, A\&A, 699, A152. doi:10.1051/0004-6361/202453347}
\bibitem[\protect\citeauthoryear{Tognelli et al.}{2021}]{2021MNRAS.501..383T} Tognelli E., Dell'Omodarme M., Valle G., Prada Moroni P.~G., Degl'Innocenti S., 2021, MNRAS, 501, 383. doi:10.1093/mnras/staa3686
\bibitem[\protect\citeauthoryear{Vieira et al.}{2022}]{2022ApJ...932...28V} Vieira K., Carraro G., Korchagin V., Lutsenko A., Girard T.~M., van Altena W., 2022, ApJ, 932, 28. doi:10.3847/1538-4357/ac6b9b
\bibitem[\protect\citeauthoryear{Vincenzo et al.}{2019}]{2019A&A...630A.125V} Vincenzo F., Miglio A., Kobayashi C., Mackereth J.~T., Montalban J., 2019, A\&A, 630, A125. doi:10.1051/0004-6361/201935886
\bibitem[\protect\citeauthoryear{Wenger et al.}{2000}]{2000A&AS..143....9W} Wenger M., Ochsenbein F., Egret D., Dubois P., Bonnarel F., Borde S., Genova F., et al., 2000, A\&AS, 143, 9. doi:10.1051/aas:2000332
\bibitem[\protect\citeauthoryear{Y{\i}ld{\i}z \& {\"O}rtel}{2021}]{2021MNRAS.504.2273Y} Y{\i}ld{\i}z M., {\"O}rtel S., 2021, MNRAS, 504, 2273. doi:10.1093/mnras/stab996
\bibitem[\protect\citeauthoryear{Y{\i}ld{\i}z}{2023}]{2023MNRAS.518.5552Y} Y{\i}ld{\i}z M., 2023, MNRAS, 518, 5552. doi:10.1093/mnras/stac3464
\bibitem[\protect\citeauthoryear{Zhang et al.}{2019}]{2019ApJS..244...43Z} Zhang J., Qian S.-B., Wu Y., Zhou X., 2019, ApJS, 244, 43. doi:10.3847/1538-4365/ab442b
\end{thebibliography}

\appendix
\section{Computation of age of red giants from the grids}
\label{sec:AppA}
\begin{figure}
    \centering
\includegraphics[width=1.3\linewidth]{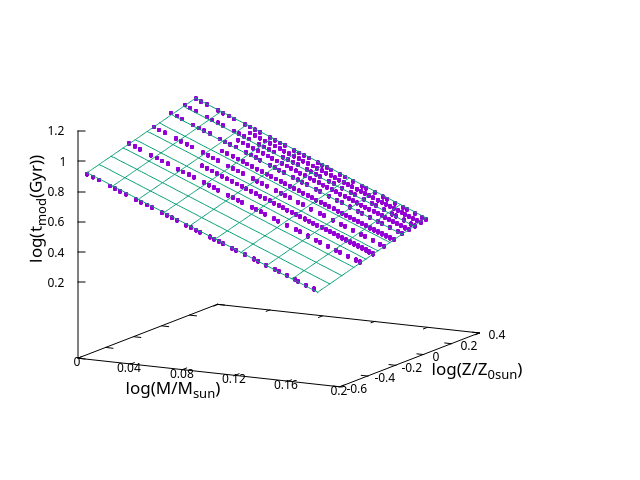}
\includegraphics[width=1.2\linewidth]{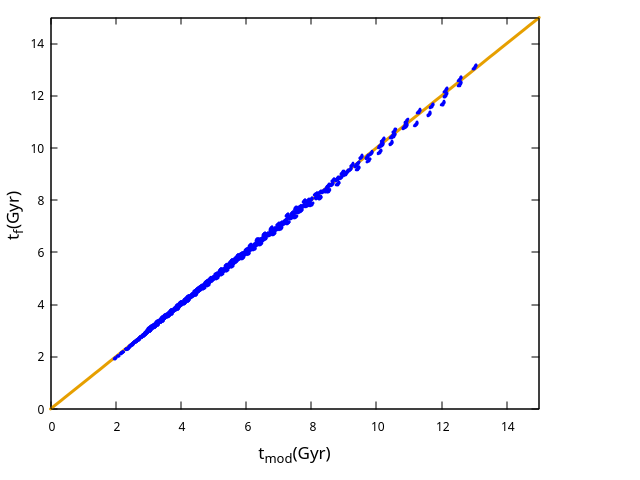}
    \caption{For the stellar grids obtained using {\small MESA }, age is plotted with respect to $M$ and $Z$ in logarithmic scales. These models are RG models with R=7.5-14 $\RS$, in accordance with the radius range of EB SLOs. The relationship between age, $M$, and $Z$ is evident and forms a plane on this graph. The fitting formula in equation (A1) is also plotted. It gives roughly $t\propto Z^{0.28}/M^{3.54}$. Taking $R$ into account, a much more accurate relationship for age is obtained (equations A2-A3). This age is plotted with respect to model age in the lower panel. The difference between the two ages is less than 3 percent.}
    \label{fig:tf_tmod}
\end{figure}
{It is well known that mass is the most influential parameter on the age of stars. However, because the evolutionary paths of stars in and around MS depend on other parameters such as $Y$, $Z$, and $\alpha$, the uncertainty in the age is generally large unless we know the star's evolutionary stage. Knowing that the SLO stars we investigate are RG stars, we can calculate their ages with much greater precision. For this purpose, we prepared a grid of RG models. In this grid, $M$ is in the range of 1.0-1.5 $\MS$, and $Z$ is in the range of 0.005-0.030. The SLO RGs we examined have radii between R=7.5 and 14 $\RS$. In Fig. \ref{fig:tf_tmod}(a), the age ($t_{\rm mod}$) of models with $R$ in this range is plotted against $M$ and $Z$. We obtain a very precise fitting formula for age ($t_{\rm f}$) in terms of $\log(M/\MS)$ and $\log(Z/Z_{0\sun})$:
\begin{align}
\label{eq:A1}
    \log(t_{\rm f}/{\rm Gyr})=&1.0503\pm0.0001+(-3.5394\pm0.0007)\log(M/\MS)+\nonumber\\ 
    &(0.2769\pm0.0002)\log(Z/Z_{0\sun}).
\end{align}
}

{A high star's mass shortens its lifetime, whereas a high $Z$ extends its lifetime. We can calculate a more precise age by including the radius:
\begin{align}
\label{eq:A2}
    &\log(t_{\rm f}/{\rm Gyr})=1.0279\pm0.0005+(-3.5440\pm0.0007)\log(M/\MS)+\nonumber \\ 
    &(0.2779\pm0.0002)\log(Z/Z_{0\sun})+(0.0222\pm0.0005)\log(R/\RS).
\end{align}
The age calculated from this formula is plotted against the model age in Fig. \ref{fig:tf_tmod}.
For RGs, we can write the age as: 
\begin{equation}
\label{eq:A3}
t_{\rm f}= 10.66 \frac{(Z/Z_{0\sun})^{0.278} R^{0.022}}{M^{3.544}}.
\end{equation}
}

{The difference between $t_{\rm f}$ and $t_{\rm mod}$ is less than 3 per cent. The ages calculated using this formula are given in Table \ref{tab:Yol12_t9}. This formula is also very convenient for calculating the age uncertainty using the quadratic or MC methods.}

\section{Notes on the individual SLO binaries}
{
\subsection{KIC 5786154 }
\label{sec:kic5786154}

Using the $Z_{\rm s}$ value of KIC 5786154 given in Table \ref{tab:EBobs}, we constructed interior models for the component stars with different $Y_0$ values by applying Method I. We found that $t_{\rm A}=t_{\rm B}=4.83$ Gyr for $Y_{0}=0.3484$. From the calibration of the radii, we obtained $\alpha_{\rm A}=2.1451$ and $\alpha_{\rm B}=2.9911$. The values of both $Y_0$ and $\alpha_{\rm B}$ are very high.  

The largest difference between the metallicity obtained from Method II ($Z_0=0.0039$) and the observational metallicity ($Z_{\rm s}=0.0117$) occurs for this system. Table \ref{tab:TgZ} lists the spectroscopic data compiled from the SIMBAD data base \citep{2000A&AS..143....9W}. The data reported by \cite{2016ApJ...832..121G}, which correspond to the highest temperature estimate, are listed in the last row.

In general, the spectroscopic measurements of this star show an interesting trend: the values of $T_{\rm eff}$, $\log g$, and [Fe/H] approximately define a plane in a three-dimensional parameter space. Using the data in Table \ref{tab:TgZ}, we derive the following relation:
$$
{\rm [Fe/H]} = -(5.65\pm0.71)\times10^{-4} T_{\rm eff} + (0.95\pm0.13)\log g .
$$

For $T_{\rm eff}\approx 4610$ K and the orbital value $\log g = 2.35$ (cgs), this relation yields ${\rm [Fe/H]} \approx -0.37$, corresponding to $Z \approx 0.0057$. This value is in much better agreement with the result obtained from Method II.

\begin{table}
    \centering
    \caption{Spectroscopic data for the KIC 5786154 system compiled from the SIMBAD data base.}
    \begin{tabular}{cllc}
        \hline
 $T_{\rm eff}$(K) & $\log g $ (cgs) & [Fe/H] & Ref.\\
        \hline
4642   &  2.579   &   -0.178  & \cite{2018ApJS..235....5Q}\\
4620   &  2.56   &   -0.160  & "\\
4610   &  2.59   &   -0.140  & "\\
4610   &  2.70   &   -0.030  & "\\
4606   &  2.60   &   -0.148  & \cite{2019ApJS..244...43Z}\\
4514   &  2.53    &   -0.13  & \cite{2016AA...594A..39F}\\
4747   &  2.60    &   -0.06  & \cite{2016ApJ...832..121G}\\
        \hline
    \end{tabular}
    \label{tab:TgZ}
\end{table}

\subsection{KIC 8410637 (Tek Ayak)}

While \cite{2016ApJ...832..121G} and earlier studies suggested that the primary component of this system may be an RC star \citep{2010ApJ...713L.187H,2016ApJ...832..121G,2016AN....337..793B}, \cite{2018MNRAS.478.4669T} confirmed that it is an RGB star.

The observational data for KIC 8410637 are taken from two different studies \citep{2013A&A...556A.138F,2018MNRAS.478.4669T}. We use the parameters given by \cite{2013A&A...556A.138F} for the interior modelling, whereas the oscillation frequencies are adopted from \cite{2018MNRAS.478.4669T}.

The primary component of KIC 8410637 can be considered a reference star (see Section \ref{sec:461}) for the asteroseismic scaling relations (equations \ref{eq:clasicsca}$-$\ref{eq:non-standardMsca}). 
For this reason, it is useful to list the model parameters obtained with Method II: $Z_0=0.01629$, $Y_0=0.2797$, $t=2.30$ Gyr, $\alpha_{\rm A}=2.1884$, $Y_{\rm s}=0.29254$, $Z_{\rm s}=0.01638$, $X_{\rm s}=0.69108$, $T_{\rm eff}=4800$ K, $R=10.69$ \RSbit, $\Gamma_{1\rm s}=1.6523$, and $\mu_{\rm s}=1.3163$. The last two quantities enter the scaling relations through $\nu_{\rm max}$.

We constructed an interior model for the primary component using the {\small MESA} astero module and computed the adiabatic oscillation frequencies for comparison with the observations. In this module, the near-surface effects are corrected using the `combined' option of \cite{2014A&A...568A.123B}. The observed and model frequencies are in excellent agreement (see Table \ref{tab:astero_freq}).

We also constructed interior models using the observational parameters derived by \cite{2018MNRAS.478.4669T}. In that study, two different $\teff$ values (6066 K for Model 841037T and 6380 K for Model 8410637TzT in Table \ref{tab:Yol12_t9}) were reported for the secondary component. For the case of $\teff=6066$ K, the difference between the two studies is $\Delta \teff =424$ K. Such a discrepancy implies that the luminosity of the secondary component in \cite{2018MNRAS.478.4669T} is underestimated.

For this low luminosity, the secondary component is found to be close to the ZAMS ($t_{\rm 9B}=0.0774$ Gyr) for $Y_0=0.2751$, whereas the age of the primary component is approximately 2.72 Gyr. Even for lower helium abundances, the age difference remains very large. 
To obtain coeval solutions with Method I, $Y_0$ must be as low as 0.1934, significantly below the primordial helium abundance. 
For Method II, agreement is obtained only for $Z_0=0.0326$, much higher than the observed metallicity reported by \cite{2018MNRAS.478.4669T}. 
These results indicate that no physically consistent solution can be obtained using the parameters from that study.

Models computed using the alternative temperature reported by \cite{2018MNRAS.478.4669T} ($\teff=6380$ K) and $Z=0.0140$ also do not yield a solution. In contrast, adopting $Z=0.0194$ from \cite{2013A&A...556A.138F} leads to convergent solutions with both Methods I and II. The resulting values of $Z_0$, $Y_0$, and age are in very good agreement between the two methods, both giving an age of approximately 2.99 Gyr (see Table \ref{tab:Yol12_t9}).

We therefore conclude that the interior models constructed using the observational constraints of \cite{2013A&A...556A.138F} are consistent with the asteroseismic inferences of \cite{2018MNRAS.478.4669T}.

\subsection{KIC 9540226 (Ayva)}

The parameters of the components of KIC 9540226 were determined in three different studies \citep{2016ApJ...832..121G,2018MNRAS.476.3729B,2018MNRAS.478.4669T}. 
These studies analysed high-quality light curves and radial velocity measurements of the system. The smallest uncertainties in the radii were reported by \cite{2016ApJ...832..121G}, and therefore we adopted their dynamical masses and radii in our modelling. 
The oscillation frequencies of the primary component and its RGB nature were established by \cite{2018MNRAS.478.4669T}.

The ages determined by Methods I and II are in very good agreement. In the MZ diagram, the primary component is located very close to the triangle base.

Interior models were also constructed using the observational constraints derived by \cite{2018MNRAS.476.3729B}. The corresponding Method I and II solutions are listed in Table \ref{tab:Yol12_t9}. The two observational data sets and the two modelling methods yield very similar ages for the system. However, the resulting $c_{\rm YZ}$ values differ significantly between the data sets. This result indicates that accurate dynamical parameters are essential for reliable determinations of $c_{\rm YZ}$. The $\Dnu$ values of the models constructed using the parameters of \cite{2018MNRAS.476.3729B} agree better with the observations, which may support $c_{\rm YZ}\approx 2$.

\subsection{KIC 7037405}

Method II yields $Z_{0}=0.0093$, in good agreement with the APOGEE DR12 metallicity ([Fe/H] $=-0.13$). The age difference between Methods I and II is approximately 6 per cent. The Method II solution appears more physically plausible because the Method I solution gives $Y_0<Y_{\rm p}$. According to Methods I and II, the ages of the system are 5.75 and 5.42 Gyr, respectively. Despite the very different chemical compositions obtained by the two methods, namely $(Z_0,Y_0)=(0.0072,0.2381)$ for Method I and $(0.0093,0.2657)$ for Method II, the derived ages are remarkably similar.

\subsection{KIC 9970396}

The parameters of the system ($Z_0$, $Y_0$, $t$, and $\alpha$ values) obtained from Methods I and II are in excellent agreement. The differences between the two solutions are negligibly small. The value of $c_{YZ}$ obtained from Method I is 1.91, very close to 2. The primary component of this system can also be used as a reference star in the asteroseismic scaling relations for RGs (see Section \ref{sec:461}).

\subsection{KIC 4054905}

Solar-like oscillations were detected in four eclipsing binaries by \cite{2021A&A...648A.113B}, including KIC 4054905. They suggested that the oscillating component is an RC star. However, \cite{2022A&A...668A..82B} later established that the SLO component is more likely an RGB star. No simultaneous solution for age and chemical composition could be obtained for this system. Only an approximate age estimate, about 9.33 Gyr, was derived from the interior model of the primary component.

\subsection{KIC 4663623}
\label{sec:4663623}

The fundamental parameters of the components of KIC 4663623 were obtained by \cite{2016ApJ...832..121G} and \cite{2021A&A...648A.113B}. The masses of the two components are nearly identical ($\approx 1.41\,\MS$). We adopted the parameters reported by \cite{2016ApJ...832..121G}.

The most remarkable feature of this system is that, despite the nearly equal masses, the primary component appears to be much more evolved than the secondary. Therefore, no common-age solution can be found. If the observational data are reliable, the primary component may have experienced mass loss.

It is possible that the primary star is an RC star that lost a significant amount of mass before reaching its current position in the HRD. We estimated the amount of mass loss by first determining the age of the secondary component. For $Z_0=0.01$, we obtained $t_{\rm B}=2.63$ Gyr. At this age, the mass required for the primary component model to satisfy the observed luminosity is approximately 1.45 $\MS$. Comparing this value with the observed mass yields an estimated mass loss of about $0.09\,\MS$.

\subsection{KIC 9246715}

Both components of this system are located in the CHeB region of the HR diagram, and their masses are very similar ($M\approx2.17\,\MS$) \citep{2016ApJ...818..108R}. Their radii are also very similar ($R\approx8.37\,\RS$).

The dynamical and asteroseismic parameters are adopted from \cite{2016ApJ...818..108R} and \cite{2019MNRAS.484..451H}. The period spacing and $\Delta\nu$ of the primary component were determined by \cite{2020A&A...639A..63G} as 150 s and 8.310 $\mu$Hz, respectively. These values indicate that the primary component is a secondary clump star.

Constructing interior models for binaries containing CHeB stars is difficult because of two major uncertainties: (a) the amount of mass lost by the evolved component and (b) the amount of mass accreted by the companion. The approximate age estimated from the interior model of the primary component is about 0.78 Gyr.

\subsection{KIC 10001167}

The dynamical and asteroseismic parameters of the system were reported by \cite{2016ApJ...832..121G} and \cite{2025A&A...699A.152T}. According to the former study, the primary component lies on the left side of the triangle in the MZ diagram, implying possible mass loss.

We first constructed interior models assuming constant mass and adopting $Z=0.0027$. A significant discrepancy was found between the component ages: the age of the lower-mass star is about 12 Gyr, whereas the SLO component gives about 7 Gyr. The discrepancy can only be removed by adopting an unrealistically low helium abundance ($Y_0=0.1518$) in Method I. Method II yields a relatively high metallicity ($Z_0=0.00921$) and an age of 19.81 Gyr, exceeding the age of the Milky Way. These extreme values indicate that this observational data set is not suitable for reliable modelling.

We also constructed interior models using the parameters derived by \cite{2025A&A...699A.152T}. Method I again yields a low helium abundance ($Y_0=0.2167$) and an age of 10.81 Gyr. Method II gives $Z_0=0.0050$ and an age of 10.08 Gyr. In this case, the age becomes lower than the age of the Milky Way. If this metallicity is close to the true value of the system, the oscillating component lies inside the triangle in the MZ diagram, implying that no mass loss is required for the observed mass of the primary component ($0.9337\,\MS$).

\subsection{KIC 7377422}

The dynamical and asteroseismic parameters of the components are taken from \cite{2016ApJ...832..121G}. We obtained solutions for the stellar parameters using both Methods I and II. The ages determined by Methods I and II are 5.79 and 6.13 Gyr, respectively. The helium abundance obtained from Method I ($Y_0=0.2868$) is relatively high for $Z_0=0.0063$, corresponding to a large value of $c_{YZ}=6.30$.

\subsection{KIC 8430105}

This system has the smallest mass ratio among the 11 EBs: $q=M_{\rm B}/M_{\rm A}=0.634$. The ages obtained from Methods I and II using the observational data of \cite{2016ApJ...832..121G} are very similar. However, the corresponding $c_{YZ}$ value obtained from Method I is very small ($c_{YZ}=0.25$).

Interior models were also constructed using the observational constraints derived by \cite{2022MNRAS.517.4187T}. For Method I, the resulting $Y_0$ is so small that $c_{\rm YZ}<0$. In contrast, Method II yields results very similar to those obtained using the parameters of \cite{2016ApJ...832..121G}.
}



\bsp	
\label{lastpage}
\end{document}